\documentclass[a4paper,UKenglish,cleveref, autoref, thm-restate,final]{lipics-v2021}
\title{Reaching Your Goal Optimally by~Playing~at~Random with no Memory} %TODO Please add
\titlerunning{Reaching Your Goal Optimally by Playing at Random with no Memory} %TODO optional, please use if title is longer than one line

\author{Benjamin Monmege}{Aix Marseille Univ, Universit\'e de Toulon,
	CNRS, LIS, Marseille,
	France}{benjamin.monmege@univ-amu.fr}{https://orcid.org/0000-0002-4717-9955}{}%

\author{Julie Parreaux}{ENS Rennes,
	France}{julie.parreaux@ens-rennes.fr}{}{}%

\author{Pierre-Alain Reynier}{Aix Marseille Univ, Universit\'e de
	Toulon, CNRS, LIS, Marseille,
	France}{pierre-alain.reynier@univ-amu.fr}{}{}

\authorrunning{B. Monmege, J. Parreaux, and P.-A. Reynier}
\Copyright{Benjamin Monmege, Julie Parreaux, and Pierre-Alain Reynier} %TODO mandatory, please use full first names. LIPIcs license is "CC-BY";  http://creativecommons.org/licenses/by/3.0/

 %TODO mandatory; please add comma-separated list of keywords

\category{} %optional, e.g. invited paper

\relatedversion{} %optional, e.g. full version hosted on arXiv, HAL, or other respository/website
\supplement{}%optional, e.g. related research data, source code, ... hosted on a repository like zenodo, figshare, GitHub, ...

\funding{Benjamin Monmege and Pierre-Alain Reynier are partly funded by ANR project Ticktac (ANR-18-CE40-0015).}%optional, to capture a funding statement, which applies to all authors. Please enter author specific funding statements as fifth argument of the \author macro.

\nolinenumbers %uncomment to disable line numbering

\hideLIPIcs  %uncomment to remove references to LIPIcs series (logo, DOI, ...), e.g. when preparing a pre-final version to be uploaded to arXiv or another public repository

\theoremstyle{definition}
\newtheorem*{assumption}{Assumption}

\usepackage[svgnames,usenames,dvipsnames]{xcolor}
\usepackage{xparse}
\usepackage[obeyFinal]{todonotes}
\usepackage{wasysym}

\usepackage[linesnumbered,ruled,vlined]{algorithm2e}

\usepackage{tikz}
\usetikzlibrary{arrows,shapes,decorations,automata,backgrounds,positioning}

\tikzstyle{PlayerMin}=[draw,circle,minimum size=7mm,inner sep=1.5pt]
\tikzstyle{target}=[circle, minimum size=1mm,inner sep=-2pt]
\tikzstyle{PlayerMax}=[draw,rectangle,minimum size=7mm,inner sep=1.5pt]
\tikzstyle{proba}=[draw,circle,minimum height=0pt,inner sep=2pt,minimum width=0pt,fill=black]
\tikzset{every loop/.style={looseness=7}, >=latex}

\renewcommand\geq{\geqslant}
\renewcommand\leq{\leqslant}
\newcommand\Z{\mathbb Z}
\newcommand\Zbar{\Z_{\infty}}
\newcommand\N{\mathbb N}
\DeclareMathOperator*{\argmin}{\arg\!\min}

\newcommand\tuple[1]{\langle #1 \rangle}

\renewcommand\epsilon{\varepsilon}

\newcommand\MinPl{\mathsf{Min}}
\newcommand\MaxPl{\mathsf{Max}}

\newcommand\vertices{V}
\newcommand\minvertices{\vertices_{\MinPl}}
\newcommand\maxvertices{\vertices_{\MaxPl}}
\newcommand\edges{E}
\newcommand\finalvertices{T}

\newcommand\edgeweights{\omega}
\newcommand\game{\mathcal G}
\newcommand\Payoff{\mathbf{P}}
\ProvideDocumentCommand{\gameEx}{o}{\IfNoValueTF{#1}{\tuple{\vertices,\edges,\edgeweights,\Payoff}}{\tuple{\vertices,\edges,\edgeweights,#1}}} 

\ProvideDocumentCommand{\RPayoff}{o}{\IfNoValueTF{#1}{\mathbf{RP}}{#1\text{-}\mathbf{RP}}}
\newcommand\arenaEx{\tuple{\maxvertices,\minvertices,\edges,\edgeweights,\finalvertices}} 

\newcommand\strategy{\sigma}
\newcommand\minstrategy{\sigma}
\newcommand\randomminstrategy{\rho}
\newcommand\maxstrategy{\tau}
\newcommand\randommaxstrategy{\chi}

\newcommand\MDP{\game^{\randomminstrategy}}
\newcommand\MC{\game^{\randomminstrategy,\randommaxstrategy}}

\newcommand\outcomes{\mathsf{Play}}

\newcommand\dStrat{\mathsf{d}\Sigma}
\newcommand\mStrat{\mathsf{m}\Sigma}

\newcommand\Value{\mathsf{Val}}
\newcommand\dValue{\mathsf{dVal}}
\newcommand\mValue{\mathsf{mVal}}
\newcommand\pValue{\Value}

\newcommand\duppervalue{\overline{\dValue}}
\newcommand\dlowervalue{\underline{\dValue}}
\newcommand\muppervalue{\overline{\mValue}}

\newcommand\puppervalue{\overline{\pValue}}
\newcommand\plowervalue{\underline{\pValue}}

\newcommand\fake{\mathsf{fake}}

\newcommand\TP{\textnormal{\textbf{TP}}}
\ProvideDocumentCommand{\SP}{o}{\IfNoValueTF{#1}{\mathbf{SP}}{#1\text{-}\mathbf{SP}}}

\newcommand\Distr[1]{\Delta(#1)}
\newcommand\support[1]{\mathop{\mathrm {supp}}(#1)}
\newcommand\Proba{\mathbb P}
\newcommand\Dirac[1]{\mathsf{Dirac}_{#1}}

\newcommand{\kpos}{Ic + \frac{w^+}{w^-}Ic + \frac{2|\dValue^{\minstrategy^n}(v_0)| + |V|W}{w^-}c + |V|}

\newcommand{\kneg}{\frac{|V|W + 2|\max(-n,~\dValue(v_0))|}{c^-_{\max}}c + |V|}
\newcommand{\I}{2(\ln (w^+ + |V|W) - \ln \epsilon + \ln 8)}

\newcommand\E{\mathbb{E}}

\begin{document}

\maketitle
	
\begin{abstract}
  Shortest-path games are two-player zero-sum games played on a graph
  equipped with integer weights. One player, that we call $\MinPl$,
  wants to reach a target set of states while minimising the total
  weight, and the other one has an antagonistic objective. This
  combination of a qualitative reachability objective and a
  quantitative total-payoff objective is one of the simplest settings 
  where $\MinPl$ needs memory (pseudo-polynomial in the weights) to
  play optimally. In this article, we aim at studying a tradeoff
  allowing $\MinPl$ to play at random, but using no memory. We show
  that $\MinPl$ can achieve the same optimal value in both cases. In
  particular, we compute a randomised memoryless $\varepsilon$-optimal
  strategy when it exists, where
  probabilities are parametrised by $\varepsilon$. We also show that
  for some games, no optimal randomised strategies exist. We then
  characterise, and decide in polynomial time, the class of games
  admitting an optimal randomised memoryless strategy.
\end{abstract}

%\newpage

\section{Introduction}
\label{sec:introduction}

Game theory is now an established model in the computer-aided design
of correct-by-construction programs. Two players, the controller and
an environment, are fighting
 one against the other in a zero-sum game
played on a graph of all possible configurations. A winning strategy
for the controller results in a correct program, while the environment
is a player modelling all uncontrollable events that the program must
face. Many possible objectives have been studied in such two-player zero-sum
games played on graphs: reachability, safety, repeated reachability,
and even all possible $\omega$-regular objectives \cite{GraTho02}.

Apart from such \emph{qualitative} objectives, more
\emph{quantitative} ones are useful in order to select a particular
strategy among all the ones that are correct with respect to a
qualitative objective. Some metrics of interest, mostly studied in the
quantitative game theory literature, are mean-payoff,
discounted-payoff, or total-payoff. All these objectives have in
common that both players have strategies using no memory or randomness
to win or play optimally \cite{GimZie04}.

Combining quantitative and qualitative objectives, enabling to select
a good strategy among the valid ones for the selected metrics, often
leads to the need of memory to play optimally. One of the simplest
combinations showing this consists in the shortest-path games combining
a reachability objective with a total-payoff quantitative objective
(studied in~\cite{KhaBor08,BGHM17} under the name of \emph{min-cost
  reachability games}). Another case of interest is the combination of
a parity qualitative objective (modelling every possible
$\omega$-regular condition), with a mean-payoff objective (aiming for
a controller of good quality in the average long-run), where
controllers need memory, and even infinite memory, to play optimally
\cite{ChaHen05}.

It is often crucial to enable randomisation in the strategies. For
instance, Nash equilibria are only ensured to exist in matrix games
(like rock-paper-scissors) when players can play at random
\cite{Nas50}. In the context of games on graphs, a player may choose,
depending on the current history, the probability distribution on the
successors. In contrast, strategies that do not use randomisation are
called \emph{deterministic} (we sometimes say \emph{pure}).

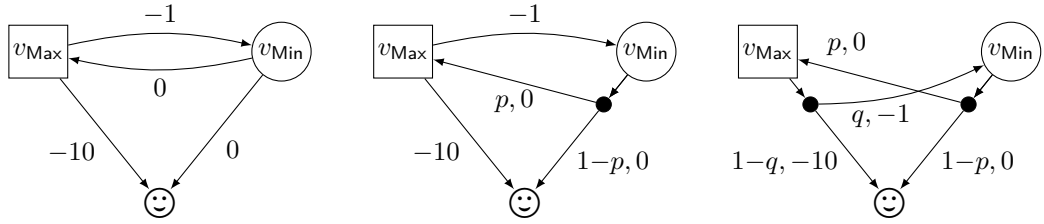
\begin{figure}[tbp]
  \centering
  \begin{tikzpicture}[xscale=.8]
    
    % Draw the states
    \node[PlayerMax] at (0, 0)  (s0){$v_{\MaxPl}$}; %{\textcolor{DarkViolet}{$-10$}};
    \node[PlayerMin] at (4, 0)  (s) {$v_{\MinPl}$}; %{\textcolor{DarkViolet}{$-10$}};
    \node[target] at (2, -2) (s1) {\LARGE $\smiley$};
    
    % Connect the states with arrows
    \draw[->]
    (s)  edge [bend left=10] node[below]{$0$} (s0) % node[very near start]{\textcolor{ForestGreen}{$p$}} (s0)
    (s0) edge [bend left=10] node[above] {$-1$} (s)
    (s0) edge [auto = right] node {$-10$} (s1)
    (s) edge [auto = left] node{$0$} (s1) %node[very near
    % start]{\textcolor{blue}{$1-p$}} (s1)
    %(s1) edge[loop right] node[right] {$0$} (s1)
    ;

    \begin{scope}[xshift=6cm]
      % Draw the states
      \node[PlayerMax] at (0, 0)  (s0){$v_{\MaxPl}$}; %{\textcolor{DarkViolet}{$-10$}};
      \node[PlayerMin] at (4, 0)  (s) {$v_{\MinPl}$}; %{\textcolor{DarkViolet}{$-10$}};
      \node[target] at (2, -2) (s1) {\LARGE $\smiley$};

      \node[draw,circle,minimum height=0pt,inner sep=2pt,minimum width=0pt,fill=black] at (3.3,-.7) (s2) {};
      
      % Connect the states with arrows
      \draw[->]
      (s)  edge (s2) % node[very near
      % start]{\textcolor{ForestGreen}{$p$}}
      % (s0)
      (s0) edge [bend left=10] node[above] {$-1$} (s)
      (s0) edge [auto = right] node {$-10$} (s1)
      (s) edge [auto = left] (s2) %node[very near
      % start]{\textcolor{blue}{$1-p$}} (s1)
      (s2) edge node[below] {$p,0$} (s0)
      (s2) edge node[below right,yshift=1mm] {$1{-}p,0$} (s1)
      %(s1) edge[loop right] node[right] {$0$} (s1)
      ;
      
    \end{scope}
    \begin{scope}[xshift=12cm]
      % Draw the states
      \node[PlayerMax] at (0, 0)  (s0){$v_{\MaxPl}$}; %{\textcolor{DarkViolet}{$-10$}};
      \node[PlayerMin] at (4, 0)  (s) {$v_{\MinPl}$}; %{\textcolor{DarkViolet}{$-10$}};
      \node[target] at (2, -2) (s1) {\LARGE $\smiley$};

      \node[proba] at (3.3,-.7) (s2) {};
      \node[proba] at (0.7,-.7) (s3) {};
      
      % Connect the states with arrows
      \draw[->]
      (s)  edge (s2) % node[very near
      % start]{\textcolor{ForestGreen}{$p$}}
      % (s0)
      (s0) edge (s3)
      (s3) edge [bend right=10,auto=right] node[very near start,yshift=1mm] {$q,-1$} (s)
      (s3) edge [auto = right] node[yshift=1mm] {$1{-}q,-10$} (s1)
      (s) edge [auto = left] (s2) %node[very near
      % start]{\textcolor{blue}{$1-p$}} (s1)
      (s2) edge[auto=right] node[very near end] {$p,0$} (s0)
      (s2) edge node[below right,yshift=1mm] {$1{-}p,0$} (s1)
      %(s1) edge[loop right] node[right] {$0$} (s1)
      
      ;
      
    \end{scope}
  \end{tikzpicture}
  \caption{On the left, a shortest-path game, where $\MinPl$ requires
    memory to play optimally. In the middle, the Markov Decision
    Process obtained when letting $\MinPl$ play at random, with a
    parametric probability $p\in(0,1)$. On the right, the Markov Chain
    obtained when $\MaxPl$ plays along a memoryless randomised
    strategy, with a parametric probability $q\in[0,1]$.}
  \label{fig:SP1}
\end{figure}
In this article, we will focus on shortest-path games, as the one
depicted on the left of \figurename~\ref{fig:SP1}. The objective of
$\MinPl$ is to reach vertex $\text{\Large\smiley}$, while minimising
the total weight. Let us consider the vertex $v_\MinPl$ as
initial. Player $\MinPl$ could reach directly $\text{\Large\smiley}$,
thus leading to a payoff of $0$. But he can also choose to go to
$v_\MaxPl$, in which case $\MaxPl$ either jumps directly
in~$\text{\Large\smiley}$ (leading to a beneficial payoff $-10$), or
comes back to $v_\MinPl$, but having already capitalised a total
payoff $-1$. We can continue this way \emph{ad libitum} until $\MinPl$
is satisfied (at least 10 times) and jumps to
$\text{\Large\smiley}$. This guarantees a value at most $-10$ for
$\MinPl$ when starting in $v_\MinPl$. Reciprocally, $\MaxPl$ can
guarantee a payoff at least $-10$ by directly jumping into
$\text{\Large\smiley}$ when she must play for the first time. Thus,
the optimal value is $-10$ when starting from~$v_\MinPl$
or~$v_\MaxPl$. However, $\MinPl$ cannot achieve this optimal value by
playing \emph{without memory} (we sometimes say \emph{positionally}),
since it either results in a total-payoff $0$ (directly going to the
target) or $\MaxPl$ has the opportunity to keep $\MinPl$ in the
negative cycle for ever, thus never reaching the target. Therefore,
$\MinPl$ needs memory to play optimally. He can do so by playing a
\emph{switching strategy}, turning in the negative cycle long enough
so that no matter how he reaches the target finally, the value he gets
as a payoff is lower than the optimal value. This strategy uses
pseudo-polynomial memory with respect to the weights of the game
graph.

In this example, such a switching strategy can be \emph{mimicked}
using randomisation only (and no memory), $\MinPl$ deciding to go to
$v_\MaxPl$ with high probability $p<1$ and to go to the target vertex
with the remaining low probability $1-p>0$ (we enforce this
probability to be positive, in order to reach the target with
probability $1$, no matter how the opponent is playing). The resulting
\emph{Markov Decision Process (MDP)} is depicted in the middle of
\figurename~\ref{fig:SP1}. The shortest path problem in such MDPs has
been thoroughly studied in~\cite{BertsekasTsitsiklis91}, where it is
proved that $\MaxPl$ does not require memory to play
optimally. Denoting by~$q$ the probability that $\MaxPl$ jumps in
$v_\MinPl$ in its memoryless strategy, we obtain the \emph{Markov
  chain (MC)} on the right of \figurename~\ref{fig:SP1}. We can
compute (see Example~\ref{example:for-intro}) the expected value in
this MC, as well as the best strategy for both players: in the
overall, the optimal value remains $-10$, even if $\MinPl$ no longer
has an optimal strategy. He rather has an $\varepsilon$-optimal
strategy, consisting in choosing $p=1-\varepsilon/10$ that ensures a
value at most $-10+\varepsilon$.

This article thus aims at studying the tradeoff between memory and
randomisation in strategies for shortest-path games. The study is only
interesting in the presence of both positive and negative weights,
since both players have optimal memoryless deterministic strategies
when the graph contains only non-negative weights~\cite{KhaBor08}. The
tradeoff between memory and randomisation has already been
investigated in many classes of games where memory is required to win
or play optimally. This is for instance the case for qualitative games
like Street or Müller games thoroughly studied (with and without
randomness in the arena) in \cite{ChaAlfHen04}. The study has been
extended to timed games \cite{ChaHenPra208} where the goal is to use
as little information as possible about the precise values of
real-time clocks. Memory or randomness is also crucial in
multi-dimensional objectives \cite{ChaRan14}: for instance, in
mean-payoff parity games, if there exists a deterministic
finite-memory winning strategy, then there exists a randomised
memoryless almost-sure winning strategy.

In contrast to previous work, we show that deterministic memory and
memoryless randomisation provide the same power to $\MinPl$. We leave
the combination of memory and randomisation for future work, as
explained in the discussion. After a presentation of the model of
shortest-path games in Section~\ref{sec:preliminaries}, we show in
Section~\ref{sec:memory2random} how the previous simulation of memory
with randomisation can be performed for all shortest-path games. The
general case is much more challenging, in particular in the presence
of positive cycles in the graph, that $\MinPl$ cannot avoid in
general. Section~\ref{sec:random2memory} shows reciprocally how to
mimic randomised strategies with memory
only. Section~\ref{sec:optimality} studies the optimality of
randomised strategies. Indeed, all shortest-path games admit an
optimal deterministic strategy for both players, but $\MinPl$ may
require memory to play optimally (even with randomisation allowed). We
thus characterises the shortest-path games in which $\MinPl$ admits an
optimal memoryless strategy, and decide this characterisation in
polynomial time.

\section{Shortest-path games: deterministic or memoryless strategies} 
\label{sec:preliminaries}

In this section, we formally introduce the shortest-path games we
consider throughout the article, as already thoroughly studied
in~\cite{BGHM17} under the name of \emph{min-cost reachability
  games}. We denote by $\Z$ the set of integers, and
$\Zbar=\Z\cup\{-\infty,+\infty\}$. For a finite set $V$, we denote by
$\Distr{V}$ the set of \emph{distributions} over $V$, that are all
mappings $\delta\colon V\to [0,1]$ such that
$\sum_{v\in V} \delta(v)=1$. The support of a distribution $\delta$ is
the set $\{v\in V\mid \delta(v)>0\}$, denoted by $\support\delta$. A
Dirac distribution is a distribution with a singleton support: the
Dirac distribution of support $\{v\}$ is denoted by $\Dirac v$. 

We consider two-player turn-based games played on weighted graphs and
denote the players by $\MaxPl$ and $\MinPl$. Formally, a
\emph{shortest-path game} (SPG) is a tuple $\arenaEx$ where
$\vertices:=\maxvertices\uplus \minvertices\uplus \finalvertices$ is a
finite set of vertices partitioned into the sets $\maxvertices$ and
$\minvertices$ of $\MaxPl$ and $\MinPl$ respectively, and a set
$\finalvertices$ of target vertices,
$\edges\subseteq \vertices\setminus\finalvertices\times \vertices$ is
a set of \emph{directed edges}, and $\edgeweights\colon \edges \to \Z$
is the \emph{weight function}, associating an integer weight with each
edge. In the drawings, $\MaxPl$ vertices are depicted by rectangles;
$\MinPl$ vertices by circles. For every vertex~$v\in\vertices$, the
set of successors of $v$ with respect to $\edges$ is denoted by
$\edges(v) = \{v'\in\vertices\mid (v,v')\in \edges\}$. Without loss of
generality, we assume that non-target vertices are deadlock-free,
i.e. for all vertices~$v\in \vertices\setminus \finalvertices$,
$\edges(v)\neq\emptyset$. Finally, throughout this article, we let
$W=\max_{(v,v')\in\edges}|\edgeweights(v,v')|$ be the greatest edge
weight (in absolute value) in the arena. A \emph{finite play} is a
finite sequence of vertices $\pi=v_0v_1\cdots v_k\in \vertices^*$ such
that for all $0\leq i<k$, $(v_i,v_{i+1})\in \edges$. Its \emph{total
  weight} is the sum $\sum_{i=0}^{k-1} \edgeweights(v_i,v_{i+1})$ of
its weights. A \emph{play} is either a finite play ending in a target
vertex, or an infinite sequence of vertices $\pi = v_0v_1\cdots$
avoiding the target such that every finite prefix $v_0\cdots v_k$,
denoted by $\pi[k]$, is a finite play.

The total-payoff of a play $\pi=v_0v_1\ldots$ is given by
$\TP(\pi)=+\infty$ if the play is infinite (and therefore avoids
$\finalvertices$), or by the total weight
$\TP(\pi)= \sum_{i=0}^{k-1} \edgeweights(v_i,v_{i+1})$ if
$\pi=v_0 v_1 \cdots v_k$ is a finite play ending in a vertex
$v_k\in\finalvertices$ (for the first time).

% the target is obtained by summing up the weights along $\pi$,
% i.e.~$\TP(\pi) = \sum_{i=0}^{k-1} \edgeweights(v_i,v_{i+1})$. The
% shortest-path payoff $\SP(\pi)$ of a play $\pi=v_0v_1\ldots$ is
% given by $\SP(\pi)=+\infty$ if the play avoids $\finalvertices$,
% i.e.~if for all $k\geq 0$, $v_k\notin\finalvertices$, and
% $\SP(\pi)=\TP(\pi[k])$ if $k$ is the least position in $\pi$ such
% that $v_k\in\finalvertices$.

A \emph{strategy} for $\MinPl$ over an arena $\game=\arenaEx$ is a
mapping
$\minstrategy\colon \vertices^* \minvertices \to \Distr{\vertices}$
such that for all sequences $\pi= v_0\cdots v_k$ with
$v_k\in \minvertices$, the support of the distribution
$\minstrategy(\pi)$ is included in $\edges(v_k)$. A play or finite
play $\pi = v_0v_1\cdots$ conforms to the strategy $\minstrategy$ if
for all~$k$ such that $v_k\in \minvertices$, we have that
$\minstrategy(\pi[k])(v_{k+1})>0$. A similar definition allows one to
define strategies
$\maxstrategy\colon \vertices^* \maxvertices \to \Distr{\vertices}$
for $\MaxPl$, and plays conforming to them.

A strategy $\strategy$ is \emph{deterministic} (or \emph{pure}) if for
all finite plays $\pi$, $\strategy(\pi)$ is a Dirac distribution: in
this case, we let $\strategy(\pi)$ denote the unique vertex in the
support of this Dirac distribution. We let $\dStrat_\MinPl$ and
$\dStrat_\MaxPl$ be the deterministic strategies of players $\MinPl$
and $\MaxPl$, respectively.
A strategy $\strategy$ is \emph{memoryless} if for all finite plays
$\pi, \pi'$, and all vertices $v\in\vertices$, we have that
$\strategy(\pi v)=\strategy(\pi' v)$ for all $v\in \vertices$. We let
$\mStrat_\MinPl$ and $\mStrat_\MaxPl$ be the memoryless strategies of
players $\MinPl$ and $\MaxPl$, respectively. To distinguish them
easily from deterministic strategies, we will denote a memoryless
strategy of $\MinPl$ using letter $\randomminstrategy$ (for
\emph{r}andom).

In this article, we focus on deterministic strategies on the one hand,
and memoryless strategies on the other hand. Even if the notion of
values that we will now introduce could be defined in a more general
setting, we prefer to give two simpler definitions in the two separate
cases, for the sake of clarity.

\subsection{Deterministic strategies}

In case of deterministic strategies, for all vertices $v$, we let
$\outcomes(v,\minstrategy,\maxstrategy)$ be the unique play conforming
to strategies $\minstrategy$ and $\maxstrategy$ of $\MinPl$ and
$\MaxPl$, respectively, and starting in~$v$. This unique play has a
payoff $\TP(\outcomes(v,\minstrategy,\maxstrategy))$. Then, we define
the value of strategies $\minstrategy$ and $\maxstrategy$ by letting
for all $v$,
\[
\dValue^{\minstrategy}(v)= \sup_{\maxstrategy'\in\dStrat_\MaxPl}
\TP(\outcomes(v,\maxstrategy',\minstrategy))
\qquad \text{ and } \qquad
  \dValue^{\maxstrategy}(v)= \inf_{\minstrategy'\in\dStrat_\MinPl}
  \TP(\outcomes(v,\maxstrategy,\minstrategy'))
\]
Finally, the game
itself has two possible values, an \emph{upper value} describing the
best $\MinPl$ can hope for, and a \emph{lower value} describing the
best $\MaxPl$ can hope for: for all vertices $v$,
\[\duppervalue(v) = \inf_{\minstrategy\in\dStrat_\MinPl}
  \dValue^{\minstrategy}(v) \qquad\text{ and } \qquad
  \dlowervalue(v) = \sup_{\maxstrategy\in\dStrat_\MaxPl}
  \dValue^{\maxstrategy}(v)
\] We may easily show that
$\dlowervalue(v)\leq \duppervalue(v)$ for all initial
vertices $v$.  In \cite[Theorem~1]{BrihayeGeeraertsHaddadMonmege2016},
shortest-path games are shown to be determined when both players use
deterministic strategies,
i.e.~$\dlowervalue(v)=\duppervalue(v)$. We thus denote
$\dValue(v)$ this common value. We say that deterministic
strategies $\minstrategy^\star$ of $\MinPl$ and $\maxstrategy^\star$
of $\MaxPl$ are optimal (respectively, $\varepsilon$-optimal for a
positive real number $\varepsilon$) if, for all vertices $v$:
$\dValue^{\minstrategy^\star}(v)=\dValue(v)$ and
$\dValue^{\maxstrategy^\star}(v)=\dValue(v)$
(respectively,
$\dValue^{\minstrategy^\star}(v)\leq
\dValue(v)+\varepsilon$ and
$\dValue^{\maxstrategy^\star}(v)\geq
\dValue(v)-\varepsilon$). % If the game is clear from the context,
% we may drop the index $\game$ from all previous notations.

\begin{example}\label{ex:needs-memory}
  The deterministic value of the game on the left of
  \figurename~\ref{fig:SP1} is described in the introduction:
  $\dValue(v_\MinPl)=\dValue(v_\MaxPl)=-10$. An optimal
  strategy for player $\MinPl$ consists in going to $v_\MaxPl$ the
  first $10$ times, and switching to the target vertex afterwards. An
  optimal strategy for player $\MaxPl$ consists in directly going
  towards the target vertex. 

  \noindent If we remove the edge from $v_\MaxPl$ to the target (of
  weight $-10$), we obtain another game in which
  $\dValue(v_\MinPl)=\dValue(v_\MaxPl)=-\infty$ since $\MinPl$ can
  decide to turn as long as he wants in the negative cycle, before
  switching to the target. There is no optimal strategy for $\MinPl$
  but a sequence of strategies guaranteeing a value as low as we want.
\end{example}

\subsection{Memoryless strategies}

Definitions above can be adapted for memoryless (randomised)
strategies. In order to keep the explanations simple, we only define
the upper value above, without relying on hypothetical determinacy
results in this context\todo{Surtout qu'on ne sait toujours pas si
  c'est vrai}. Once we fix a memoryless (randomised) strategy
$\randomminstrategy\in\mStrat_\MinPl$, we obtain a \emph{Markov
  decision process} (MDP) where the other player must still choose how
to react. An MDP is a tuple $\langle V, A, P\rangle$ where $V$ is a
set of vertices, $A$ is a set of actions, and
$P\colon V\times A\to \Distr V$ is a partial function mapping to some
pair of vertices and actions a distribution of probabilities over the
successor vertices. In our context, we let $\MDP$ be the
MDP with the same set $V$ of vertices as $\game$, actions
$A=V\cup\{\bot\}$ being either successor vertices of the game or an
additional action $\bot$ denoting the random choice of
$\randomminstrategy$, and a probability distribution $P$ defined by:
\begin{itemize}
\item if $v\in \maxvertices$, $P(v,v')$ is only defined if
  $(v,v')\in\edges$ in which case $P(v,v')=\Dirac {v'}$, and
  $P(v,\bot)$ is also undefined;
\item if $v\in \minvertices$, $P(v,\bot)=\randomminstrategy(v)$, and
  $P(v,v')$ is undefined for all $v'\in\vertices$.
\end{itemize}
In drawings of MDPs (and also of Markov chains, later), we show
weights as trivially transferred from the game graph.

\begin{example}
  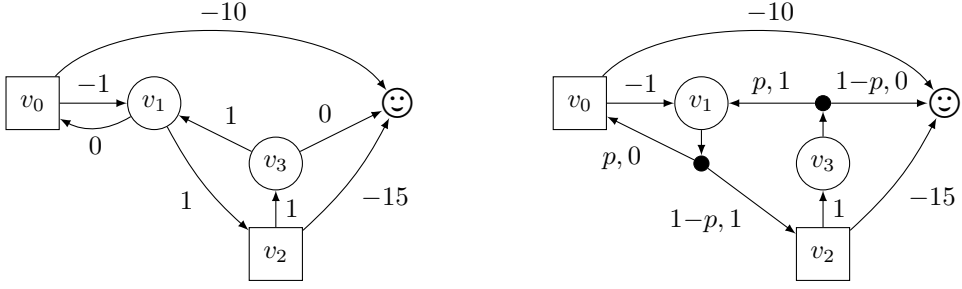
\begin{figure}[tbp]
    \centering
    \begin{tikzpicture}[xscale=.8,yscale=.8]
      % Draw the states
      \node[PlayerMax] (s0){$v_0$}; % {\textcolor{DarkViolet}{$-10$}};
      \node[PlayerMin] at (2, 0) (s1){$v_1$}; % {\textcolor{DarkViolet}{$-10$}};
      \node[PlayerMax] at (4, -2.5) (s2){$v_2$};% {\textcolor{DarkViolet}{$-8$}};
      \node[PlayerMin] at (4, -1) (s3){$v_3$};%{\textcolor{DarkViolet}{$-9$}};
      \node[target] at (6, 0) (s4) {\LARGE $\smiley$};
      
      % Connect the states with arrows
      \draw[->]
      (s1)  edge [auto = left, bend left] node {$0$} (s0)
      (s0) edge [auto = left] node {$-1$} (s1)
      (s0) edge [auto = left, bend left=50,looseness=.7] node {$-10$} (s4)
      (s1) edge [auto = right,bend right=10] node {$1$} (s2)
      (s2) edge [auto = right] node {$1$} (s3)
      (s3) edge [auto = right] node {$1$} (s1)
      (s3) edge [auto = left] node {$0$} (s4)
      (s2) edge [auto = right,bend right=10] node {$-15$} (s4)
      ;

      \begin{scope}[xshift=9cm]
        % Draw the states
        \node[PlayerMax] (s0){$v_0$}; % {\textcolor{DarkViolet}{$-10$}};
        \node[PlayerMin] at (2, 0) (s1){$v_1$}; % {\textcolor{DarkViolet}{$-10$}};
        \node[PlayerMax] at (4, -2.5) (s2){$v_2$};% {\textcolor{DarkViolet}{$-8$}};
        \node[PlayerMin] at (4, -1) (s3){$v_3$};%{\textcolor{DarkViolet}{$-9$}};
        \node[target] at (6, 0) (s4) {\LARGE $\smiley$};

        \node[proba] at (2, -1) (p1) {};
        \node[proba] at (4, 0) (p3) {};
        
        % Connect the states with arrows
        \draw[->]
        (s1)  edge (p1)
        (p1) edge [auto = left] node {$p,0$} (s0)
        (s0) edge [auto = left] node {$-1$} (s1)
        (s0) edge [auto = left, bend left=50,looseness=.7] node {$-10$} (s4)
        (p1) edge [auto = right] node {$1{-}p,1$} (s2)
        (s2) edge [auto = right] node {$1$} (s3)
        (s3) edge (p3)
        (p3) edge [auto = right] node {$p,1$} (s1)
        (p3) edge [auto = left] node[xshift=-1mm] {$1{-}p,0$} (s4)
        (s2) edge [auto = right,bend right=10] node {$-15$} (s4)
        ;
      \end{scope}
    \end{tikzpicture}
    \caption{On the left, a more complex example of shortest-path
      game. On the right, the MDP associated with a randomised
      strategy of $\MinPl$ with a parametric probability
      $p\in (0,1)$.}
    \label{fig:SP2}
  \end{figure}
  In \figurename~\ref{fig:SP1}, a shortest-path game is presented on
  the left, with the MDP in the middle obtained by picking as a
  memoryless strategy for $\MinPl$ the one choosing to go
  to~$v_\MaxPl$ with probability $p\in(0,1)$ and to the target vertex
  with probability $1-p$. Another more complex example is given in
  \figurename~\ref{fig:SP2} where the memoryless strategy for $\MinPl$
  consists, in vertex~$v_1$, to choose successor $v_0$ with
  probability $p\in(0,1)$ and successor $v_2$ with probability~$1-p$,
  and in vertex $v_3$, to choose successor $v_1$ with the same
  probability $p$ and the target vertex with probability $1-p$.
\end{example}

In such an MDP, when player $\MaxPl$ has chosen her strategy, there
will remain no ``choices'' to make, and we will thus end up in a
\emph{Markov chain}. A Markov chain (\emph{MC}) is a tuple
$\mathcal M= \langle V, P\rangle$ where $V$ is a set of vertices, and
$P\colon V\to \Distr V$ associates to each vertex a distribution of
probabilities over the successor vertices. In our context, for all
memoryless strategies $\randommaxstrategy\in\mStrat_\MaxPl$, we
let $\MC$ the MC obtained from the MDP $\MDP$ by following strategy
$\randommaxstrategy$ and action $\bot$. Formally, it consists of the
same set $V$ of vertices as $\game$, and mapping $P$ associating to a
vertex $v\in\minvertices$, $P(v)=\randomminstrategy(v)$ and to a
vertex $v\in\maxvertices$, $P(v)=\randommaxstrategy(v)$.

\begin{example}
  On the right of \figurename~\ref{fig:SP1} is depicted the MC
  obtained when $\MaxPl$ decides to go to $v_\MinPl$ with probability
  $q\in[0,1]$ and to the target vertex with probability $1-q$.
\end{example}

When starting in a given initial vertex $v$, we let
$\Proba^{\randomminstrategy,\randommaxstrategy}_{v}$ denote the
induced probability measure over the sets of paths in the MC $\MC$ (as
before, $\game$ is made implicit in the notation). A \emph{property}
is any measurable subset of finite or infinite paths in the MC with
respect to the standard cylindrical sigma-algebra. For instance, we
denote by
$\Proba^{\randomminstrategy,\randommaxstrategy}_{v} (\diamond
T)$ the probability of the set of plays that reach the target set
$T\subseteq V$ of vertices. Given a random variable $X$ over the
infinite paths in the MC, we let
$\E^{\randomminstrategy,\randommaxstrategy}_{v}(X)$ be the
expectation of $X$ with respect to the probability measure
$\Proba^{\randomminstrategy,\randommaxstrategy}_{v}$. Therefore,
$\E^{\randomminstrategy,\randommaxstrategy}_{v}(\TP)$ is the
expected weight of a path in the MC, weights being the ones taken from
$\game$.

The objective of $\MaxPl$ is to maximise the payoff in the MDP
$\MDP$. We therefore define the value of strategy
$\randomminstrategy$ of $\MinPl$ as the best case scenario for $\MaxPl$:
\[\mValue^{\randomminstrategy}(v)=
  \sup_{\randommaxstrategy\in\mStrat_\MaxPl}
  \E^{\randomminstrategy,\randommaxstrategy}_{v}(\TP)\] 
  %This
%definition only makes sense \todo{makes sense -> is a real number?} (otherwise it is $+\infty$) if
By \cite[Section~10.5.1]{BaierKatoen}\todo{C'est uniquement avec poids
  positifs ici, mais la preuve est identique dans le cas général}, the
value $\mValue^{\randomminstrategy}(v)$ is finite if and only if
$\Proba^{\randomminstrategy,\randommaxstrategy}_{v} (\diamond T)=1$
for all $\randommaxstrategy$, i.e.~if strategy $\randomminstrategy$
ensures the reachability of a target vertex with probability 1, no
matter how the opponent plays. In this case, letting $P$ be the
probability mapping defining the MC $\MC$, the vector
$(\E^{\randomminstrategy,\randommaxstrategy}_{v}(\TP))_{v\in
  \vertices}$ is the only solution of the system of equations
\begin{equation}
  \label{eq:Bellman}
  \E^{\randomminstrategy,\randommaxstrategy}_{v}(\TP) =
  \begin{cases}
    0 & \text{if } v\in \finalvertices\\
    \sum_{v'\in E(v)} P(v,v') \times
    (\edgeweights(v,v')+\E^{\randomminstrategy,\randommaxstrategy}_{v'}(\TP))
    & \text{if } v\notin\finalvertices
  \end{cases}
\end{equation}

Since $\MinPl$ wants
to minimise the shortest-path payoff, we finally define the memoryless
upper value as
\[\muppervalue(v) =
  \inf_{\randomminstrategy\in\mStrat_\MinPl}
  \mValue^{\randomminstrategy}(v)\] Once again, we say that a
memoryless strategy $\randomminstrategy$ is optimal (respectively,
$\varepsilon$-optimal for a positive real number $\varepsilon$) if
$\mValue^{\randomminstrategy}(v)=\muppervalue(v)$
(respectively,
$\mValue^{\randomminstrategy}(v)\leq
\muppervalue(v)+\varepsilon$). With respect to player
$\MaxPl$, we only consider optimality and $\varepsilon$-optimality in
the MDP $\MDP$.

\begin{example}\label{example:for-intro}
  For the game of \figurename~\ref{fig:SP1}, we let $\minstrategy$ and
  $\maxstrategy$ the memoryless strategies that result in the MC on
  the right. Letting
  $x=\E_{v_\MinPl}^{\randomminstrategy,\randommaxstrategy}(\TP)$ and
  $y=\E_{v_\MaxPl}^{\randomminstrategy,\randommaxstrategy}(\TP)$, the
  system \eqref{eq:Bellman} rewrites~as
  $x = (1-p)\times 0 + p \times y$ and $y =
    q\times (-1 + x) + (1-q) \times (-10)$.
  We thus have $x=p(9q-10)/(1-pq)$. Two cases happen, depending on the
  value of $p$: if $p<9/10$, then $\MaxPl$ maximises $x$ by choosing
  $q=1$, while she chooses $q=0$ when $p\geq 9/10$. In all cases,
  player $\MaxPl$ will therefore play deterministically: if $p<9/10$,
  the expected payoff from $v_\MinPl$ will then be
  $\mValue^{\randomminstrategy}(v_\MinPl)=-p/(1-p)$; if
  $p\geq 9/10$, it will be
  $\mValue^{\randomminstrategy}(v_\MinPl)=-10p$. This value is
  always greater than the optimum $-10$ that $\MinPl$ were able to
  achieve with memory, since we must keep $1-p>0$ to ensure reaching
  the target with probability $1$. We thus obtain
  $\muppervalue(v_\MinPl)= \muppervalue(v_\MaxPl)=-10$
  as before. There are no optimal strategies for $\MinPl$, but an
  $\varepsilon$-optimal one consisting in choosing probability
  $p\geq 1-\varepsilon/10$.
\end{example}

The fact that $\MaxPl$ can play optimally with a deterministic
strategy in the MDP $\game^{\randomminstrategy}$ is not specific to
this example. Indeed, in an MDP $\game^{\randomminstrategy}$ such that
$\Proba^{\randomminstrategy,\randommaxstrategy}_{v} (\diamond T)=1$
for all $\randommaxstrategy$, $\MaxPl$ cannot avoid reaching the
target: she must then ensure the most expensive play
possible. Considering the MDP $\tilde{\game}^{\randomminstrategy}$
obtained by multiplying all the weights in the graph by $-1$, the
objective of $\MaxPl$ becomes a shortest-path objective. We can then
deduce from \cite{BertsekasTsitsiklis91} that she has an optimal
deterministic memoryless strategy: the same applies in the original
MDP $\game^{\randomminstrategy}$.

\begin{proposition}%[Optimal strategy for $\MaxPl$]
  \label{prop:strat_opti_Eve}
  In the MDP $\game^{\randomminstrategy}$ such that
  $\Proba^{\randomminstrategy,\randommaxstrategy}_{v} (\diamond T)=1$
  for all $\randommaxstrategy$, $\MaxPl$ has an optimal deterministic
  memoryless strategy.
\end{proposition}

\subsection{Contribution}

Our contribution consists in showing that optimal values are the same
when restricting both players to memoryless or deterministic
strategies:
\begin{theorem}\label{thm:main}
  For all games $\game$ with a shortest-path objective, for all
  vertices $v$, we have $\dValue(v)=\muppervalue(v)$. 
\end{theorem}

We show this theorem in the two next sections by a simulation of
deterministic strategies with memoryless ones, and vice versa. We
start here by ruling out the case of values $+\infty$. Indeed,
$\dValue(v)=+\infty$ signifies that $\MinPl$ is not able to reach a
target vertex from $v$ with deterministic strategies. This also
implies that $\MinPl$ has no memoryless randomised strategies to
ensure reaching the target with probability $1$, and thus
$\muppervalue(v)=+\infty$. Reciprocally, if~$\muppervalue(v)=+\infty$,
then $\MinPl$ has no memoryless strategies to reach the target with
probability $1$ (since this is the only reason for having a value
$+\infty$). Since reachability is a purely qualitative objective, and
the game graph does not contain probabilities, $\MinPl$ cannot use
memory in order to guarantee reaching the target: therefore, this also
means that $\dValue(v)=+\infty$. In the end, we have shown that
$\dValue(v)=+\infty$ if and only if $\muppervalue(v)=+\infty$. We thus
remove every such vertex from now on, which does not change the values
of other vertices in the game.

\begin{assumption}
  From now on, all games $\game$ with a shortest-path objective are
  such that~$\dValue(v)$ and~$\muppervalue(v)$ are
  different from $+\infty$, for all vertices $v$.
\end{assumption}

\section{Simulating deterministic strategies with memoryless
  strategies}
\label{sec:memory2random}

Towards proving Theorem~\ref{thm:main}, we show in this section that,
for all shortest-path games $\game=\gameEx$ (where no values are
$+\infty$) and vertices $v\in \vertices$,
$\muppervalue(v) \leq \dValue(v)$. This is done by considering the
\emph{switching strategies} originated from
\cite{BrihayeGeeraertsHaddadMonmege2016}, which are a particular kind
of deterministic strategies: they are optimal from vertices of finite
value, and they can get a value as low as wanted from vertices of
value $-\infty$. A switching strategy
$\minstrategy=\langle \minstrategy_1,\minstrategy_2,\alpha\rangle$ is
described by two deterministic memoryless strategies $\minstrategy_1$
and $\minstrategy_2$, as well as a switching parameter~$\alpha$. The
strategy $\minstrategy$ consists in playing along $\minstrategy_1$,
until eventually switching to~$\minstrategy_2$ when the length of the
current finite play is greater than $\alpha$.  Strategy
$\minstrategy_2$ is thus any \emph{attractor strategy} ensuring that
plays reach the target set of vertices: it can be computed via a
classical attractor computation. Strategy $\minstrategy_1$ is chosen
so that every cyclic finite play $v_0v_1\ldots v_kv_0$ conforming to
$\minstrategy_1$ has a negative total weight: this is called an
\emph{NC-strategy} (for \emph{negative-cycle-strategy}) in
\cite{BrihayeGeeraertsHaddadMonmege2016}. The \emph{fake-value} of
$\minstrategy_1$ from a vertex $v_0$ is defined by
$\fake^{\minstrategy_1}(v_0)=\sup\{\TP(v_0v_1\cdots v_k)\mid v_k\in
\finalvertices, v_0v_1\cdots v_k \text{ conforming to }
\minstrategy_1\}$, letting $\sup\emptyset=-\infty$: it consists of
only considering plays conforming $\minstrategy_1$ that reach the
target. Strategy $\minstrategy_1$ is said to be \emph{fake-optimal} if
$\fake^{\minstrategy_1}(v)\leq \dValue(v)$ for all vertices~$v$: in
this case, if a play from $v$ conforms to $\minstrategy_1$ (or
$\minstrategy$ before the switch happens) and reaches the target set
of vertices, it has a weight at most $\dValue(v)$.

\begin{proposition}[\cite{BrihayeGeeraertsHaddadMonmege2016}]\label{prop:switching}
  There exists a fake-optimal NC-strategy $\minstrategy_1$. Moreover,
  for all such fake-optimal NC-strategies $\minstrategy_1$, for all
  attractor strategies $\minstrategy_2$, and for all $n\in \N$, the
  switching parameter $\alpha=(2W(|V|-1)+n)|V|+1$ defines a switching
  strategy
  $\minstrategy=\langle \minstrategy_1,\minstrategy_2,\alpha\rangle$
  with a value $\dValue^\minstrategy(v) \leq \max(-n,\dValue(v))$,
  from all initial vertices $v\in\vertices$.
\end{proposition}

In particular, if $\dValue(v)$ is finite, for $n$ large enough,
the switching strategy is optimal. If $\dValue(v)=-\infty$
however, the sequence $(\minstrategy^n)_{n\in\N}$ of strategies, each
with a different parameter~$n$, has a value that tends to $-\infty$.

\begin{example}
  For all $n \in \N$, let
  $\minstrategy = (\minstrategy_1, \minstrategy_2, \alpha)$ the
  switching strategy described above. In Figure~\ref{fig:SP1}, we have
  $\minstrategy_1(v_{\MinPl}) = v_{\MaxPl}$,
  $\minstrategy_2(v_{\MinPl}) = \text{\Large\smiley}$ and
  $\alpha = 3(40+n)+1$. In Figure~\ref{fig:SP2}, $\minstrategy_1$
  chooses $v_0$ from $v_1$ and $v_1$ from $v_3$, $\minstrategy_2$
  chooses $v_2$ from $v_1$ and {\Large\smiley} from $v_3$ and
  $\alpha = 5(60+n)+1$, for all $n \in \N$.
\end{example}

\noindent \textbf{Definition of a memoryless (randomised) strategy.}
%To show that $\muppervalue(v) \leq \dValue(v)$ for all vertices
%$v\in \vertices$, 
Let $n\in \N$,
we consider the switching strategy
$\minstrategy=\langle\minstrategy_1,\minstrategy_2,\alpha\rangle$
described before, of value
$\dValue^{\minstrategy}(v)\leq \max(-n,\dValue(v))$, and simulate it
with a memoryless (randomised) strategy for $\MinPl$, denoted
$\randomminstrategy_p$, with a parametrised probability
$p\in(0,1)$. 
This new strategy is a probabilistic superposition of the
two memoryless deterministic strategies $\minstrategy_1$ and
$\minstrategy_2$. 

Formally, we define $\randomminstrategy_{p}$ on each strongly
connected components (SCC) of the graph according to the presence of a
negative cycle. In an SCC that does not contain negative cycles, for
each vertex $v \in \minvertices$ of the SCC, we let
$\randomminstrategy_{p}(v) = \Dirac{\minstrategy_1(v)}$: player
$\MinPl$ chooses to play the first strategy $\minstrategy_1$ of the
switching strategy, thus looking for a negative cycle in the next SCCs
(in topological order) if any. In an SCC that contains a negative
cycle, for each vertex $v \in \minvertices$ of the SCC, we let
$\randomminstrategy_{p}(v)$ be the distribution of support
$\{\minstrategy_1(v), \minstrategy_2(v)\}$ that chooses
$\minstrategy_1(v)$ with probability $p$ and $\minstrategy_2(v)$ with
probability $1 - p$, except if $\minstrategy_1(v)=\minstrategy_2(v)$
in which case we choose it with probability 1. Note that MDPs in
Figures~\ref{fig:SP1} and~\ref{fig:SP2} are obtained by applying this
strategy $\randomminstrategy_{p}$.

%Strategy $\randomminstrategy_{p}$ will be such that,
%for all $\varepsilon>0$, there exists $p \in (0, 1)$ where
%$\mValue^{\randomminstrategy_{p}}(v) \leq
%\dValue^{\minstrategy}(v)+\varepsilon$. In particular, 

We fix some vertex $v_0\in V$.
In the rest of this section, we prove the following result:

\begin{restatable}{proposition}{calculAvec} \label{prop:calcul_p-avec}
  For $\epsilon$ small enough and $p$ close enough to $1$,
  \mbox{$\mValue^{\randomminstrategy_{p},\maxstrategy}(v_0) \leq
    \dValue^{\minstrategy}(v_0) + \epsilon$}.
\end{restatable}

%\begin{proposition}\label{prop:eps-optimal}
%For all $\varepsilon>0$, there exists $p \in (0, 1)$ such that for all vertices
%$v\in V$, 
%$\mValue^{\randomminstrategy_{p}}(v) \leq
%\dValue^{\minstrategy}(v)+\varepsilon$.
%\end{proposition}

This entails the expected result. Indeed, if $\dValue(v_0)\in\Z$, we get
(with $n=|\dValue(v_0)|$) that
$\mValue^{\randomminstrategy_{p}}(v_0) \leq \dValue(v_0)+\varepsilon$, and
thus $\muppervalue(v_0) \leq \dValue(v_0)$ since this holds for all
$\varepsilon>0$. Otherwise, $\dValue(v_0)=-\infty$, and letting $n$ tend
towards $+\infty$, we also get $\muppervalue(v_0) =-\infty$.

We first prove that $\randomminstrategy_{p}$ is one of the strategies of
$\MinPl$ that guarantee to reach the target with probability $1$ in
the MDP $\game^{\randomminstrategy_{p}}$ no matter how $\MaxPl$ reacts.
\begin{proposition}\label{prop:target-proba1}
  For all strategies $\randommaxstrategy\in\mStrat_\MaxPl$,
  $\Proba^{\randomminstrategy_{p}, \randommaxstrategy}_{v_0}(\diamond~T) = 1$. 
\end{proposition}
\begin{proof}
  Recall that we designed our graph games so that target vertices are
  the only deadlocks. Thus, by using the characterisation of
  \cite[Lemma 10.111]{BaierKatoen},
  $\min_{\randommaxstrategy\in\mStrat_\MaxPl}\Proba^{\randomminstrategy_{p},
    \randommaxstrategy}_{v_0}(\diamond~T) = 1$ if and only if for all
  $\randommaxstrategy\in\mStrat_\MaxPl$, all bottom SCCs of the MC
  $\game^{\randomminstrategy_{p},\randommaxstrategy}$ (the ones from which we
  cannot exit) consist in a unique target vertex. Suppose in the
  contrary that $\MaxPl$ has a memoryless strategy $\randommaxstrategy$ such
  that the MC $\game^{\randomminstrategy_{p},\randommaxstrategy}$ contains a
  bottom SCC $\mathcal C$ with no target vertices.

  If all vertices of $\mathcal C$ belong to $\MaxPl$, then they all
  have a successor in $\mathcal C$ and therefore there also exists a
  deterministic memoryless strategy $\maxstrategy'$ for which all
  vertices $v\in\mathcal C$ are such that
  $\dValue^{\maxstrategy'}(v)=+\infty$, and thus $\dValue(v)=+\infty$:
  this contradicts our hypothesis that all vertices have a
  deterministic value different from $+\infty$.

  Otherwise, for all vertices $v\in \minvertices\cap\mathcal C$, since
  $\mathcal C$ is a bottom SCC of
  $\game^{\randomminstrategy_{p},\randommaxstrategy}$, the
  distribution $\randomminstrategy_{p}(v)$ has its support included in
  $\mathcal C$. If $\mathcal C$ is included in a SCC of $\game$ with
  no negative cycles,
  $\support{\randomminstrategy_{p}(v)}=\{\minstrategy_1(v)\}$: playing
  $\minstrategy_1(v)$ in $\mathcal C$ will end up in a cycle (since
  there are no deadlocks) that must be negative, by the hypothesis on
  $\minstrategy_1$, which is impossible. Thus, $\mathcal C$ must be
  included in an SCC of $\game$ with a negative cycle. Then,
  $\support{\randomminstrategy_{p}(v)}=\{\minstrategy_1(v),
  \minstrategy_2(v)\}\subseteq \mathcal C$, and in particular the
  attractor strategy is not able to reach a target vertex: playing the
  deterministic switching strategy $\minstrategy$ will result in not
  reaching a target vertex either, so that $\dValue(v)=+\infty$ for
  $v\in \minvertices\cap\mathcal C$, which also contradicts our
  hypothesis.
\end{proof}

We can therefore apply Proposition~\ref{prop:strat_opti_Eve}. This
result is very helpful since it allows us to only consider
deterministic memoryless strategies $\maxstrategy$ to compute
$\mValue^{\randomminstrategy_{p}}(v_0)= \sup_{\maxstrategy}
\mValue^{\randomminstrategy_{p},\maxstrategy}(v_0)$, for all initial
vertices $v_0$. We thus consider such a strategy $\maxstrategy$ and we
now show that
$\mValue^{\randomminstrategy_{p},\maxstrategy}(v_0) \leq
\dValue^{\minstrategy}(v)+\epsilon$ whenever $p<1$ is close enough to
$1$ (in function of $\epsilon>0$). By gathering the finite number of
lower bounds about $p$, for all 
%initial vertices $v_0$ and
deterministic memoryless strategies of $\MaxPl$ (there are a finite
number of such), we obtain a lower bound for $p$ such that
$\mValue^{\randomminstrategy_{p}}(v_0) \leq
\dValue^{\minstrategy}(v_0)+\varepsilon$,
% for all $v$, 
as expected to
prove Proposition~\ref{prop:calcul_p-avec}.%one inequality of Theorem~\ref{thm:main}.

% We will compute $p$ in function of $\epsilon$ such that
% $\randomminstrategy_{p}$ guarantee $\dValue(v)$ as value for all $v$. By
% the Proposition~\ref{prop:strat_opti_Eve}, deterministic memoryless
% strategies for $\MaxPl$ suffice to evaluate
% $\mValue^{\randomminstrategy_{p}}$. Let $v_0$ be an initial vertex, we
% calculate $p$ such that, for all
% $\maxstrategy \in \dStrat_{\MaxPl} \cup \mStrat_\MaxPl$,
% $\E^{\randomminstrategy_{p}, \maxstrategy}_{v_0}\left[\diamond T\right]
% \leq \max(-n, \dValue(v_0) + \epsilon)$. This calculation is done
% according to the cycles present in the game's arena.

The case where the whole game graph does not contain any negative
cycles is easy. In this case, $\randomminstrategy_{p}$ chooses the
strategy $\minstrategy_1$ with probability $1$, by definition since no
SCC contain a negative cycle (this is the only reason why we defined
$\randomminstrategy_p$ as it is, for such SCCs): a play from initial
vertex $v_0$ conforming to $\randomminstrategy_{p}$ is thus conforming
to $\minstrategy_1$. Since the graph contains no negative cycles and
all cycles conforming to $\minstrategy_1$ must be negative, all plays
from $v_0$ conforming to $\minstrategy_1$ reach the target set of
vertices, with a total payoff at most
$\dValue^{\minstrategy}(v_0)$. This single play has probability $1$ in
the MC $\game^{\randomminstrategy_{p}, \maxstrategy}$, thus
$\E^{\randomminstrategy_{p}, \maxstrategy}_{v_0} (\TP) \leq
\dValue^{\minstrategy}(v_0)$, which proves that
$\mValue^{\randomminstrategy_{p}}(v) \leq \dValue^{\minstrategy}(v_0)$
as expected.

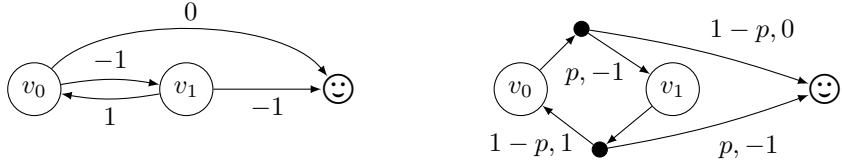
\begin{figure}[tbp]
  \centering
  \begin{tikzpicture}[xscale=.8,yscale=.8,node distance=2cm]
    % Draw the states
    \node[PlayerMin] (s0){$v_0$};
    \node[PlayerMin,right of=s0] (s1){$v_1$};
    % \node[PlayerMax] at (1.8, -4) (s2){$v_2$};  
    % \node[PlayerMax] at (3, -3.05) (s3){$v_3$};
    \node[target,right of=s1] (s4) {\LARGE $\smiley$};
    
    % Connect the states with arrows
    \draw[->]
    (s0)  edge [auto = left, bend left=10] node {$-1$} (s1)
    (s1) edge [auto = left, bend left=10] node {$1$} (s0)
    (s0) edge [auto = left, bend left=50,looseness=.7] node {$0$} (s4)
    (s1) edge [auto = right] node {$-1$} (s4)
    %	(s2) edge [auto = right] node {$0$} (s3)
    % (s3) edge [auto = right] node {$0$} (s4)
    ;

    \begin{scope}[xshift=8cm]
      \node[PlayerMin] (s0){$v_0$};
      \node[PlayerMin,right of=s0] (s1){$v_1$};
      % \node[PlayerMax] at (1.8, -4) (s2){$v_2$};  
      % \node[PlayerMax] at (3, -3.05) (s3){$v_3$};
      \node[target,right of=s1] (s4) {\LARGE $\smiley$};

      \node[proba] at (1, 1) (p0) {};
      \node[proba] at (1.3, -1) (p1) {};

      % Connect the states with arrows
      \draw[->]
      (s0) edge (p0)
      (s1) edge (p1)
      (p0)  edge [auto = right] node[xshift=2mm] {$p,-1$} (s1)
      (p1) edge [auto = left] node[xshift=2mm] {$1-p,1$} (s0)
      (p0) edge [auto = left,bend left=5] node {$1-p,0$} (s4)
      (p1) edge [auto = right,bend right=5] node {$p,-1$} (s4)
      %	(s2) edge [auto = right] node {$0$} (s3)
      % (s3) edge [auto = right] node {$0$} (s4)
      ;
    \end{scope}
  \end{tikzpicture}
  \caption{On the left, a game graph with no negative cycles
    where $\randomminstrategy_p$ is optimal. The MC obtained
    when playing a different randomised memoryless strategy. }
  \label{fig:SP3}
\end{figure}

\begin{example} \label{ex:SCCNonNeg} If the definition of
  $\randomminstrategy_p$ would not distinguish the SCCs with no
  negative cycles from the other SCCs, we would not have the
  optimality of $\randomminstrategy_p$ as shown before. Indeed,
  consider the game graph on the left of Figure~\ref{fig:SP3}, which
  has no negative cycles. We have $\dValue(v_0)=-2$ and
  $\dValue(v_1)=-1$. As a switching strategy, we can choose
  $\minstrategy_1(v_0) = v_1$,
  $\minstrategy_1(v_1) = \text{\Large\smiley}$,
  $\minstrategy_2(v_0) = \text{\Large\smiley}$ and
  $\minstrategy_2(v_1) = v_0$. Then, $\randomminstrategy_p$ is equal
  to $\minstrategy_1$ (and thus independent of~$p$), and
  $\mValue^{\randomminstrategy_p}(v_0)=-2$ and
  $\mValue^{\randomminstrategy_p}(v_1)=-1$. However, if we would have
  chosen to still mix $\minstrategy_1$ and $\minstrategy_2$, we would
  obtain a strategy $\randomminstrategy'_p$, and the MC on the right
  of Figure~\ref{fig:SP3}. Then, we get
  $\mValue^{\randomminstrategy'_p}(v_0)=-2p^2/(1-p(1-p))$ and
  $\mValue^{\randomminstrategy'_p}(v_1)=(p^2-3p+1)/(1-p(1-p))$ whose
  limits are $-2$ and $-1$ respectively, when $p$ tends to 1. This
  strategy $\randomminstrategy'_p$ would then still be
  $\epsilon$-optimal for $p$ close enough to $1$. 
  % Suppose that $\randomminstrategy'_p$ an another memoryless
  % randomised strategy based on the probabilistic combination of
  % $\minstrategy_1$ and $\minstrategy_2$ where the definition of
  % $\randomminstrategy'_p$ is the same for all SCC. In particular,
  % $\support(\randomminstrategy'_p(v_0)) = \{\text{\Large\smiley}\}$
  % $\support(\randomminstrategy'_p(v_1)) = \{v_0, v_2\}$ where
  % $\randomminstrategy'_p(v_1)(v_0) = p$ and
  % $\randomminstrategy'_p(v_1)(v_2) = 1-p$.
  % 
  % For all $\randommaxstrategy$ a memoryless strategy for
  % $\MaxPl$. Letting
  % $x = \E^{\randomminstrategy'_p, \randommaxstrategy}_{v_0}$ and
  % $y = \E^{\randomminstrategy'_p, \randommaxstrategy}_{v_1}$, the
  % Bellman's equations give
  % \[x = p \times (-1 + y) \qquad \text{ and } \qquad y =
  %   -1 \times p + (1-p)\times (1 + x)
  % \]
  % To solve the system, we obtain $x = (-p^2 - 2p + 1)/p > -2$
  % if $p < 1$.
\end{example}

Now, suppose that the graph game contains negative cycles. We let
$c>0$ be the maximal size of an elementary cycle (that visits a vertex
at most once) in $\game$, $w^->0$ be the
opposite of the maximal weight of an elementary negative cycle in $\game$, and
$w^+\geq 0$ be the maximal weight of an
elementary non-negative cycle in $\game$ (or $0$ if such cycle does
not exist).

\begin{example}
  In the graph of Figure~\ref{fig:SP1}, we have $c = 2$, $w^- = 1$,
  and $w^+=0$ (since there is no non-negative cycles). In the game
  graph of Figure~\ref{fig:SP2}, we have $c = 3$, $w^- = 1$, and
  $w^+=3$.
\end{example}

The difficulty initiates from the possible presence of non-negative
cycles too. Indeed, when applying the switching strategy
$\minstrategy$, all cycles conforming to $\minstrategy_1$ have a
negative weight. This is no longer true with the probabilistic
superposition $\randomminstrategy_{p}$, as can be seen in the example of
Figure~\ref{fig:SP2}. Finding an adequate lower-bound for $p$ requires
to estimate $\E^{\randomminstrategy_{p}, \maxstrategy}_{v_0} (\TP)$, by
controlling the weight and probability of non-negative cycles,
balancing them with the ones of negative cycles. The crucial argument
comes from the definition of the superposition $\randomminstrategy_{p}$: 

\begin{lemma}%[Characterisation of non-negative cycles]
  \label{lem:cyclesPositifs}
  All cycles in $\game^{\randomminstrategy_{p}, \maxstrategy}$ of
  non-negative total weight contain at least one edge of probability
  $1-p$.
\end{lemma}
\begin{proof}
  Suppose on the contrary that all edges have probability $p$ or $1$,
  then the cycle is conforming to strategy $\minstrategy_1$, and has
  therefore a negative weight.
\end{proof}

\begin{proof}[Proof of Proposition~\ref{prop:calcul_p-avec}]
  We prove that
  $\mValue^{\randomminstrategy_{p},\maxstrategy}(v_0) \leq
  \dValue^{\minstrategy}(v_0) + \epsilon$ as $\MaxPl$ has 
  an optimal deterministic memoryless strategy against 
  $\minstrategy_p$. It is done by partitioning the
  set $\Pi$ of plays starting in $v_0$, conforming to
  $\randomminstrategy_{p}$ and $\maxstrategy$, and reaching the target
  set of vertices, into subsets $\Pi_{i,\ell}$ according to the number
  $i$ of edges of probability~$1-p$ they go through, and their length
  $\ell$ (we always have $i\leq \ell$). The partition is depicted in
  Figure~\ref{fig:partition}:
  \begin{itemize}
  	\item $\Pi_{0, \N}$, depicted in yellow, contains all plays with no
  	edges of probability $1-p$;
  	\item $\Pi_{> 0, \geq L}$, depicted in blue, contains all plays with
  	$i\geq 1$ edges of probability $1-p$, and a length of
  	at least
  	\[
  	L(i) = ia + b \qquad \text{with} \quad a = \left\lceil c\left(1 +
  	\frac{w^+}{w^-}\right)\right\rceil \quad \text{
  		and } \quad b = \frac{|\dValue^{\minstrategy}(v_0)| + |V|W+w^-}{w^-}c 
  	+ |V|
  	\]
  	\item $\Pi_{> 0, < L}$, depicted in red, is the rest of
  	the plays, i.e.~plays with $i \geq 1$ edges of
  	probability $1-p$ and a length less than $L(i)$. We
  	also let $\Pi_{i, <L(i)}$ be the set of plays with
  	$i\geq 1$ edges of probability $1-p$, and a length of
  	at most $L(i)$, so that $\Pi_{> 0, < L}$ is the union
  	of all such sets.
  \end{itemize}

\begin{figure}[tbp]
	\centering
	\begin{tikzpicture}[xscale=.7,yscale=.3]
	\draw[fill=DarkBlue,draw=none] (0,0) -- (9.6, 9.6) -- (0,9.6) -- (0,0);

	\draw[fill=red, draw=none] (0,0) -- (0,3.5) -- (5.5,9.6) -- (9.6,9.6);
	
	\draw[fill=yellow!70!orange,rounded corners,draw=none] (-.15,0) rectangle (.15,9.6);
	\draw[fill=white,draw=none] (-.2,9.6) rectangle (.2,10.6);

	\draw[fill=white,draw=none] (8,8) rectangle (10,10);
	
	\draw[->,thick] (0,0) -- (8.5,0) node[right](){$i$};
	
	\draw[->,thick] (0,0) -- (0,10) node[above](){$\ell$};
	\node at (6,10.25)(){$L(i)$};
	
	\node at (-.7,5) () {$\Pi_{0,\N}$};
	\node[white] at (2.25,8) () {$\Pi_{>0,\geq L}$};
	\node[white] at (4,5.6) () {$\Pi_{> 0, < L}$};
	\end{tikzpicture}
	\caption{Partition of plays $\Pi$.}
	\label{fig:partition}
\end{figure}
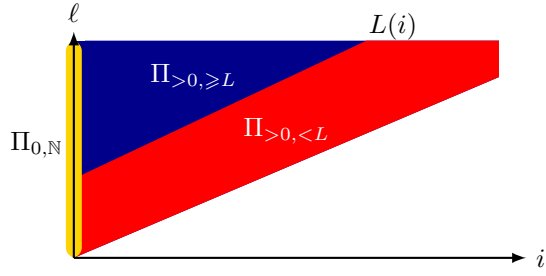

\noindent Partitioning the plays allows us to carefully control non-negative
cycles: plays with a large enough length can compensate for the presence of 
non-negative cycles and thus obtain a favorable weight 
($< \dValue^\minstrategy(v_0)$). 

We let $\gamma_{0, \N}$ (respectively, $\gamma_{> 0, \geq L}$ and
$\gamma_{> 0, < L}$) be the expectation
$\E_{v_0}^{\randomminstrategy_{p}, \maxstrategy}(\TP)$ restricted to
plays in $\Pi_{0, \N}$ (respectively, $\Pi_{> 0, \geq L}$ and 
$\Pi_{> 0, < L}$). By linearity of expectation,
\begin{equation}
\mValue^{\randomminstrategy_{p},\maxstrategy}(v_0)=
\E_{v_0}^{\randomminstrategy_{p}, 
	\maxstrategy}(\TP) = \gamma_{0, \N} + 
\gamma_{> 0, \geq L} + \gamma_{> 0, < L}
\label{eq:expectation}
\end{equation}
We thus control separately the three terms of \eqref{eq:expectation} to
obtain
$\mValue^{\randomminstrategy_{p},\maxstrategy}(v_0) \leq
\dValue^{\minstrategy}(v_0) + \epsilon$.

First, we control the weight of a play with depending on the
number of transitions $1-p$ it goes through.  Let $\pi$ be a
play in $\Pi_{i, \ell}$, with $1 \leq i$ and $\ell \geq i$: it
goes through $i$ edges of probability $1-p$. By
Lemma~\ref{lem:cyclesPositifs}, it contains at most $i$
elementary cycles of non-negative cumulated weight (at most
$w^+$). The total length of these cycles is at most $ic$. Once
we have removed these cycles from the play, it remains a play
of length at least $\ell-ic$. By a repeated pumping argument,
it still contains at least
$\left\lfloor\frac{\ell-ic-|V|}{c}\right\rfloor$ elementary
cycles, that all have a negative cumulated weight (at most
$-w^-$). The remaining part, once removed the last negative
cycles it contains, has length at most $|V|$, and thus a total
payoff at most $|V|W$. In summary the total payoff of
every play in $\Pi_{i, \ell}$ is at most
\begin{equation}
iw^+ + \left\lfloor\frac{\ell-ic-|V|}{c}\right\rfloor(-w^-) + |V|W
\label{eq:weight}
\end{equation}
Now, we control each term of~\eqref{eq:expectation}.

\paragraph*{Red zone is such that $\gamma_{>0, <L} \leq \epsilon/2$.}
Let $\pi$ be a play in $\Pi_{i, \ell}$, with $i\geq 1$ and
$\ell < L(i)$. By~\eqref{eq:weight}, its total payoff is at
most
\begin{displaymath}
iw^+ + \left\lfloor\frac{\ell-ic-|V|}{c}\right\rfloor(-w^-) + |V|W 
\leq iw^+ + |V|W
\end{displaymath}
So, we can decompose the expectation $\gamma_{>0, < L}$
as follows:
\begin{align}
\gamma_{>0, < L} 
&= \sum_{\pi \in \Pi_{> 0, < L}} \TP(\pi)
\Proba_{v_0}^{\randomminstrategy_p, \maxstrategy}
(\pi) = \sum_{i = 1}^{+\infty}\sum_{\pi \in \Pi_{i, < L(i)}} \TP(\pi)
\Proba_{v_0}^{\randomminstrategy_p, \maxstrategy}
(\pi)\nonumber\\
& \leq
\sum_{i = 1}^{+\infty} (iw^+ + |V|W)
\Proba_{v_0}^{\randomminstrategy_p, \maxstrategy}
(\Pi_{i, < L(i)}) 
\label{eq:red_expectation1}
\end{align}
Moreover, the probability of a play in $\Pi_{i, < L(i)}$,
given by the $i$ edges of probability $(1-p)$ and the
$\ell -i$ edges with a probability bounded by $1$, is at most
$(1-p)^i$. Since the number of plays in $\Pi_{i, < L(i)}$ is
bounded by $2^{L(i)}$ (for each of the at most $L(i)$ steps,
$\MinPl$ has at most 2 choices in its distribution, while
$\MaxPl$ plays a deterministic strategy), we
have\footnote{This is the novelty of this version, the
	argument given in the original version being false here.}
\begin{equation}
\Proba_{v_0}^{\randomminstrategy_p, \maxstrategy}(\Pi_{i, < L(i)}) \leq 
(1 - p)^i 2^{L(i)}
\label{eq:red_proba}
\end{equation}
We rewrite~\eqref{eq:red_expectation1} as 
\begin{align}
\gamma_{>0, < L} 
&\leq \sum_{i = 1}^{+\infty} (|V|W+iw^+) (1 - p)^i 2^{L(i)}=
\sum_{i = 1}^{+\infty} (|V|W+iw^+) (1 - p)^i
2^{ai+b}\nonumber \\ 
&\leq |V|W2^b \sum_{i = 1}^{+\infty} ((1 - p)2^a)^i
+ w^+2^b \sum_{i = 1}^{+\infty} i ((1 - p)2^a)^i \nonumber
\end{align}
these sums converging as soon as we consider $p \geq 1 -
\frac{1}{2^a}$. We finally obtain
\begin{align*}
\gamma_{>0, < L} 
&\leq |V|W2^b \frac{2^a(1-p)}{1 - 2^a(1-p)}
+ w^+2^b  \frac{2^a(1-p)}{(1 - 2^a(1-p))^2}
\end{align*}
We consider a stronger assumption on $p$, namely that
$p \geq 1 - \frac{1}{2^{a + 1}}$. Then, we know that $1 \leq
\frac{1}{1 - 2^a(1 - p)} \leq 2$, so that 
we rewrite the previous inequality as 
\begin{displaymath}
\gamma_{>0, < L} 
\leq |V|W2^{b + a + 1}(1-p) + w^+2^{b + a + 2} (1-p)
\end{displaymath}
By choosing $p$ such that
\begin{displaymath}
p \geq 1 - \frac{\epsilon}{2(|V|W2^{b + a + 1} + w^+2^{b + a + 2})}
\end{displaymath}
we obtain as desired $\gamma_{>0, <L} 
\leq\varepsilon/2$. 

\paragraph*{Yellow and blue zones are such that
	$\gamma_{0, \N} + \gamma_{>0, \geq L}\leq
	\dValue^\minstrategy(v_0) + \epsilon/2$.}
We first upper-bound the total payoffs of all plays of these
two zones.  On the one hand, all plays of $\Pi_{0, \N}$ reach
the target without edges of probability $1-p$, i.e.~by
conforming to~$\minstrategy_1$. By fake-optimality of
$\minstrategy_1$, their total payoff is upper-bounded by
$\dValue^{\minstrategy}(v_0)$. 
On the other hand, by~\eqref{eq:weight}, all plays $\pi$ of
$\Pi_{i, \ell}$, with $0 \leq i < I$ and $\ell \geq L$, have a
total payoff at most
\begin{align*}
\TP(\pi) & \leq iw^+ + \left\lfloor\frac{\ell-ic-|V|}{c}\right\rfloor (-w^-) + |V|W\\
&\leq iw^+ + \left(\frac{L - ic - |V|}{c}-1\right) (-w^-) + |V|W \\
&= iw^+ + \frac{ai + \frac{|\dValue^{\minstrategy}(v_0)| + |V|W+w^-}{w^-}c +
	|V| - ic - |V| - c}{c} (-w^-) + |V|W \\
&= iw^+ + \frac{ai + \frac{|\dValue^{\minstrategy}(v_0)| + |V|W}{w^-}c 
	- ic}{c} (-w^-) + |V|W \\
&\leq iw^+ + \left(i\left(1 + \frac{w^+}{w^-}\right) + 
\frac{|\dValue^{\minstrategy}(v_0)| + |V|W}{w^-} - i\right) (-w^-) + |V|W \\
&= iw^+ -iw^+ -|\dValue^{\minstrategy}(v_0)| -|V|W + |V|W \\
&= -|\dValue^{\minstrategy}(v_0)|\\
&\leq \dValue^\minstrategy(v_0)
\end{align*}
Therefore, all plays in the yellow and blue zones have a
payoff bounded by $\dValue^\minstrategy(v_0)$. This implies
\begin{align*}
\gamma_{0, \N} + \gamma_{> 0, \geq L}
&\leq \sum_{\pi \in \Pi_{0, \N}} \dValue^\minstrategy(v_0) 
\Proba_{v_0}^{\randomminstrategy_p, \maxstrategy}(\pi) +
\sum_{\pi \in \Pi_{> 0, \geq L}} \dValue^\minstrategy(v_0)
\Proba_{v_0}^{\randomminstrategy_p, \maxstrategy}(\pi) \\
&= \dValue^\minstrategy(v_0) 
\Proba_{v_0}^{\randomminstrategy_p, \maxstrategy}\left(\Pi_{0, \N} \cup 
\Pi_{> 0, \geq L}\right)
\end{align*}
Depending on the sign of $\dValue^\minstrategy(v_0)$, we can
conclude.
\begin{itemize}
	\item If $\dValue^\minstrategy(v_0) \geq 0$, then
	upper-bounding the probability
	$\Proba_{v_0}^{\randomminstrategy_p,
		\maxstrategy}\left(\Pi_{0, \N} \cup \Pi_{> 0, \geq
		L}\right)$ by $1$, suffices to get
	$\gamma_{0, \N} + \gamma_{> 0, \geq L(i)} \leq
	\dValue^\minstrategy(v_0)$. 
	
	\item If $\dValue^\minstrategy(v_0) < 0$, then, by the bound~\eqref{eq:red_proba}
	found for the red zone, we have
	\begin{align*}
	\Proba_{v_0}^{\randomminstrategy_p, \maxstrategy}\left(\Pi_{0, \N} \cup 
	\Pi_{> 0, \geq L}\right) 
	&= 1 - \Proba_{v_0}^{\randomminstrategy_p, \maxstrategy}
	\left(\Pi_{> 0, < L}\right)\\
	&\geq 1- \sum_{i=1}^{\infty}(1 - p)^i 2^{ai+b}\\
	&=1-\frac{2^{a+b}(1-p)}{1-(1-p)2^a}\\
	&\geq 1 - 2^{a + b+1}(1-p)  \qquad (\text{since }
	1/(1-(1-p)2^a) \leq 2)
	\end{align*}
	This allows us to obtain
	\begin{displaymath}
	\gamma_{0, \N} + \gamma_{> 0, \geq L} \leq \dValue^\minstrategy(v_0)
	(1 - 2^{a + b + 1}(1-p))
	\end{displaymath}
	In case, we have moreover
	\begin{displaymath}
	p \geq 1 - \frac{\epsilon}{2^{a + b + 2}|\dValue^\minstrategy(v_0)|}
	\end{displaymath}
	we finally obtain
	$\gamma_{0, \N} + \gamma_{> 0, \geq L} \leq
	\dValue^\minstrategy(v_0) + \frac{\epsilon}{2}$ as expected.
\end{itemize}

\paragraph*{Lower bound over $p$}

If we gather all the lower bounds over $p$ that we need in the proof,
we get that:
\begin{itemize}
	\item if
	$\dValue^{\minstrategy}(v_0)\geq 0$, we must have
	\[p\geq \max\left(1 - \frac{1}{2^{a + 1}}, 
	1 - \frac{\epsilon}{2(|V|W2^{b + a + 1} + w^+2^{b + a + 2})} \right) \]
	\item if $\dValue^{\minstrategy}(v_0)< 0$, we must have
	\[p\geq \max\left(1 - \frac{1}{2^{a + 1}}, 
	1 - \frac{\epsilon}{2(|V|W2^{b + a + 1} + w^+2^{b + a + 2})}, 
	1 - \frac{\epsilon}{2^{a + b + 2}|\dValue^\minstrategy(v_0)|}
	\right)\]
\end{itemize}
with $\epsilon$ small enough so that this bound is less than $1$.
\end{proof}

This ends the proof that for all vertices $v$,
$\muppervalue(v) \leq \dValue(v)$. Let us illustrate the computation
of the lower-bound on probability $p$ of the memoryless strategy
$\randomminstrategy_{p}$ in the previously studied examples.

\begin{example} 
  For the game in Figure~\ref{fig:SP1}, with initial vertex
  $v_{\MinPl}$, we have $a = 2$ and $b = 45$. For~$\epsilon=0.1$,
  the lower-bound on $p$ is then
  $q = 0.999999999999999994$, which gives a value
  $\mValue^{\randomminstrategy_{p}}(v_{\MinPl}) = -10p = -9.99999999999999994$. 
  For the game in Figure~\ref{fig:SP2}, with initial vertex $v_2$, we
  have $a = 12$ and $b = 257$. For $\epsilon=0.1$, the lower-bound on $p$ is then
  $q = 1 - \frac{0.1}{81 \times 2^{271}}\approx 1$, which gives a value
  $\mValue^{\randomminstrategy_{p}}(v_{2}) \approx -8$.
  We see that the lower-bound are correct, even if they could
  certainly be made coarser.
\end{example}

\section{Simulating memoryless strategies with deterministic
  strategies}
\label{sec:random2memory}

To finish the proof of Theorem~\ref{thm:main}, we will show that
$\dValue(v)\leq \muppervalue(v)$, for all vertices $v$. For a given
memoryless strategy $\randomminstrategy$ ensuring that $\MinPl$ reaches
the target set $\finalvertices$ with probability~$1$, we build a
deterministic strategy $\minstrategy$ which guarantees a value
$\dValue^\minstrategy(v)\leq\mValue^{\randomminstrategy}(v)$ from
vertex $v$. Then, as in the previous section, if $\muppervalue(v)$ is
finite, for an $\epsilon$-optimal memoryless strategy
$\randomminstrategy$, we get a deterministic strategy such that
$\dValue^\minstrategy(v) \leq \muppervalue(v)+\epsilon$, and thus
$\dValue(v)\leq \muppervalue(v)+\epsilon$. We can conclude since this
holds for all $\epsilon>0$. In case $\muppervalue(v)=-\infty$, if
$\randomminstrategy$ guarantees a value at most $-n$ with $n\in \N$,
then so does the deterministic strategy $\minstrategy$, which also
ensures that $\dValue(v)=-\infty$.

We fix a memoryless strategy $\randomminstrategy$, and an initial
vertex $v_0$. The first attempt to build a deterministic strategy
$\sigma$ such that
$\dValue^\minstrategy(v) \leq \muppervalue(v)+\epsilon$ would be to
use classical techniques of finite-memory strategies, for instance in
Street or Müller games: for instance, to ensure the visit of two
vertices $v_1$ and $v_2$ infinitely often during an infinite play (to
win a Müller game with winning objective $\{v_1,v_2\}$), we would try
to reach $v_1$ with a first memoryless strategy, and then reach $v_2$
with another memoryless strategy, before switching again to reach
$v_1$ again, etc.

\begin{example}\label{example:optimality-intro}
  Let us try this technique on the shortest-path game of
  \figurename~\ref{fig:SP1}. We consider as a starting point the
  memoryless strategy $\randomminstrategy$ such that
  $\randomminstrategy(v_\MinPl)=\delta$ with
  $\delta(\text{\Large\smiley})=2/3$ and $\delta(v_\MaxPl)=1/3$ (this
  is the case $p=1/3$ in the MDP on the middle of
  \figurename~\ref{fig:SP1}). As seen in
  Example~\ref{example:for-intro}, this strategy has value
  $\mValue^\randomminstrategy(v_\MinPl)=-1/2$ et
  $\mValue^\randomminstrategy(v_\MaxPl)=-3/2$. Naively, we could try
  to mimic the distribution $\delta$ by using memory as follows: when
  in $v_\MinPl$, go to $\text{\Large\smiley}$ two thirds of the time
  and to $v_\MinPl$ one third of the time. Moreover, we would naively
  try to follow first the choice with greatest probability. In this
  case, the strategy $\minstrategy$ would first choose to go to
  $\text{\Large\smiley}$, thus stopping immediately the play. We thus
  get $\dValue^{\minstrategy}(v_\MinPl)=0>-1/2+\varepsilon$ as soon as
  $\varepsilon<1/2$.
\end{example}

The main reason why this naive approach fails is that the plays are
essentially finite in shortest-path games. We thus cannot delay the
choices and must carefully play as soon as the play starts. Instead,
our solution is to define a switching strategy
$\minstrategy=\langle \minstrategy_1,\minstrategy_2,\alpha\rangle$,
with~$\minstrategy_2$ any attractor strategy, and
$\alpha=\max(0,|V|W-\mValue^{\rho}(v_0))\times |V|+1$.

\begin{example}[Example~\ref{example:optimality-intro} continued]
  In the game of \figurename~\ref{fig:SP1}, the attractor strategy is
  $\minstrategy_2(v_\MinPl)=\text{\Large\smiley}$. We then choose
  $\minstrategy_1(v_\MinPl)$ so as to minimise the immediate reward
  obtained by playing one turn and then getting the value ensured by
  $\randomminstrategy$:
  \[\minstrategy_1(v_\MinPl)=\textstyle{\argmin_{v'\in \{v_\MaxPl,
        \text{\Large\smiley}\}}} \left[w(v,v')+
      \mValue^{\randomminstrategy}(v')\right] = v_\MaxPl\] For an
  appropriate choice of $\alpha$, we thus recover the optimal
  switching strategy for this game.
\end{example}

In the rest of this section, we will detail how to define strategy
$\minstrategy_1$ in general so as to obtain the following property:
\begin{proposition}
  \label{prop:sigmaM}
  The switching strategy
  $\minstrategy=\langle \minstrategy_1,\minstrategy_2,\alpha\rangle$
  built from the memoryless (randomised) strategy $\randomminstrategy$
  satisfies
  $\dValue^{\minstrategy}(v_0) \leq \mValue^{\randomminstrategy}(v_0)$.
\end{proposition}

The construction of $\minstrategy_1$ is split in two parts. First, we
restrict the possibilities for $\minstrategy_1(v)$ to a subset
$\widetilde E(v)$ of $\support{\randomminstrategy(v)}$
in~\eqref{eq:Etilde}: with respect to
Example~\ref{example:optimality-intro}, this will forbid the use of
edge $(v_\MinPl, \text{\Large\smiley})$ in particular. The definition
of $\minstrategy_1(v)$ is then given later in~\eqref{eq:sigma1}.

We restrict our attention to edges present in the MDP $\MDP$, and for
each vertex $v\in\minvertices$, we let
\begin{equation}
\widetilde E(v) = \argmin_{v' \in \support{\randomminstrategy(v)}} \left[w(v,v')+
  \mValue^{\randomminstrategy}(v')\right]\label{eq:Etilde}
\end{equation} be the successors of $v$ that
minimise the expected value at horizon $1$. We let $\widetilde{\mathcal G}$ be the
game obtained from $\game$ by removing all edges $(v,v')$
from a vertex $v\in \minvertices$ such that $v'\notin \widetilde E(v)$.
\begin{lemma}\label{lem:non-positive}
  $(i)$ Each finite play of $\widetilde{\mathcal G}$ from a vertex $v$
  has a total payoff at most $\mValue^\randomminstrategy(v)$. $(ii)$
  Each cycle in the game $\widetilde{\mathcal G}$ has a non-positive
  total weight.
\end{lemma}
\begin{proof}
  We prove the property $(i)$ on finite plays $\pi$ of
  $\widetilde{\mathcal G}$ by induction on the length of~$\pi$, for
  all initial vertices $v$. If $\pi$ has length $0$, this means that
  $v\in\finalvertices$, in which case
  $\TP(\pi) = 0 = \mValue^{\randomminstrategy}(v)$. Consider then a
  play $\pi=v\pi'$ of length at least 1, with $\pi'$ starting from
  $v'$, so that $\TP(\pi)=\edgeweights(v,v')+\TP(\pi')$. By induction
  hypothesis, $\TP(\pi')\leq \mValue^{\randomminstrategy}(v')$, so
  that
  $\TP(\pi)\leq \edgeweights(v,v')+\mValue^{\randomminstrategy}(v')$.
  
  Suppose first that $v\in\maxvertices$. By
  Proposition~\ref{prop:strat_opti_Eve}, we know that $\MaxPl$ can play
  optimally in the MDP $\MDP$ with a deterministic and memoryless
  strategy. For each possible deterministic and memoryless strategy
  $\maxstrategy$ of $\MaxPl$, we have
  $\mValue^{\randomminstrategy}(u) \geq
  \E^{\randomminstrategy,\maxstrategy}_u(\TP)$ for all
  $u\in\maxvertices$, and by the system~\eqref{eq:Bellman} of equations,
  letting $u'=\maxstrategy(u)$,
  $\E^{\randomminstrategy,\maxstrategy}_u(\TP) = \edgeweights(u, u') +
  \E^{\randomminstrategy,\maxstrategy}_{u'}(\TP)$. We thus know that
  $\mValue^{\randomminstrategy}(u) \geq \edgeweights(u, u') +
  \E^{\randomminstrategy,\maxstrategy}_{u'}(\TP)$. By taking a maximum
  over all deterministic and memoryless strategies $\maxstrategy$ of
  $\MaxPl$, Proposition~\ref{prop:strat_opti_Eve} ensures that
  \begin{equation}
    \label{eq:inequality-Max}
    \forall u\in \maxvertices\quad \forall u'\in E(u)\qquad
    \mValue^{\randomminstrategy}(u) \geq \edgeweights(u, u') +
    \mValue^{\randomminstrategy}(u')
  \end{equation}
  In particular,
  $\mValue^{\randomminstrategy}(v) \geq \edgeweights(v, v') +
  \mValue^{\randomminstrategy}(v')\geq \TP(\pi)$.

  If $v\in\minvertices$, then $v'\in \widetilde E(v)$ so that
  $\edgeweights(v,v')+\mValue^{\randomminstrategy}(v')$ is minimum
  over all possible successors
  $v'\in\support{\randomminstrategy(v)}$. The system
  \eqref{eq:Bellman} of equations implies that, for an optimal
  strategy~$\randommaxstrategy$ of $\MaxPl$,
  \begin{align}
    \mValue^{\randomminstrategy}(v) 
    &=
      \E^{\randomminstrategy,\randommaxstrategy}_v(\TP) = \sum_{v''\in
      E(v)} P(v,v'')\times
      (\edgeweights(v,v'')+\E^{\randomminstrategy,\randommaxstrategy}_{v''}(\TP))
      \nonumber\\
    &    = \sum_{v''\in
      \support{\randomminstrategy(v)}} P(v,v'')\times
      (\edgeweights(v,v'')+\mValue^{\randomminstrategy}(v'')) \geq
      \edgeweights(v,v')+\mValue^{\randomminstrategy}(v')\label{eq:final-inequality}
  \end{align}
  so that we also get $\mValue^{\randomminstrategy}(v) \geq \TP(\pi)$.

  \medskip We then prove the property $(ii)$ on cycles. Consider thus
  a cycle $v_1v_2\cdots v_k v_1$ of $\widetilde{\mathcal G}$, and let
  $\edgeweights_1=\edgeweights(v_1,v_2),
  \edgeweights_2=\edgeweights(v_2,v_3), \dots,
  \edgeweights_k=\edgeweights(v_k,v_1)$ be the sequence of weights of
  edges. We also let $v_{k+1}=v_1$. We show that
  $\edgeweights_1+\edgeweights_2+\cdots +\edgeweights_k\leq 0$. Let
  $i\in\{1,2,\ldots,k\}$. If $v_i\in\maxvertices$, by
  \eqref{eq:inequality-Max},
  $\mValue^\randomminstrategy(v_i)\geq
  \edgeweights_i+\mValue^\randomminstrategy(v_{i+1})$. If
  $v_i\in\minvertices$, by the reasoning applied above in
  \eqref{eq:final-inequality}, we also know that
  $\mValue^\randomminstrategy(v_i)\geq
  \edgeweights_i+\mValue^\randomminstrategy(v_{i+1})$. By summing all
  the inequalities above, we get
  \[\sum_{i=1}^k \mValue^\randomminstrategy(v_i)\geq 
    \sum_{i=1}^k \edgeweights_i+ \sum_{i=1}^k
    \mValue^\randomminstrategy(v_{i}) \qquad \text{i.e.} \quad
    \edgeweights_1+\edgeweights_2+\cdots +\edgeweights_k\leq 0
    \qedhere\]
\end{proof}

\begin{example}
  Consider again the game graph on the left of
  \figurename~\ref{fig:SP3}, and the memoryless strategy
  $\randomminstrategy'_p$ giving rise to the MDP/MC on the right of
  \figurename~\ref{fig:SP3}. Recall that
  $\mValue^{\randomminstrategy'_p}(v_0)=-2p^2/(1-p(1-p))$ and
  $\mValue^{\randomminstrategy'_p}(v_1)=(p^2-3p+1)/(1-p(1-p))$. Consider
  $p$ close enough to $1$ so that
  $\mValue^{\randomminstrategy'_p}(v_0)\leq -3/2$ and
  $\mValue^{\randomminstrategy'_p}(v_1)\leq -1/2$.  Then, we have
  $\widetilde E(v_0) = \{v_1\}$ and
  $\widetilde E(v_1)=\{\text{\Large\smiley}\}$. The corresponding game
  graph $\widetilde {\mathcal G}$ contains only edges $(v_0,v_1)$ and
  $(v_1, \text{\Large\smiley})$, and thus no cycles. The unique finite
  play from vertex $v_0$ has total-payoff
  $-2\leq \mValue^{\randomminstrategy'_p}(v_0)$. In particular, the
  only possible memoryless deterministic strategy $\minstrategy_1$ in
  $\widetilde {\mathcal G}$ is optimal in $\game$.
\end{example}

For each vertex $v$ in the game, we let $d(v)$ be the distance (number
of steps) of $v$ to the target given by an attractor computation to
the target in $\MDP$ (notice that this may be different from the
distance given in the whole game graph, since some edges are taken
with probability~$0$ in~$\randomminstrategy$, but still $d(v)<+\infty$
since $\randomminstrategy$ ensures to reach $\finalvertices$ with
probability $1$). We then let, for all vertices~$v\in\minvertices$,
\begin{equation}
  \label{eq:sigma1}
  \minstrategy_1(v) = \argmin_{v'\in \widetilde E(v)}d (v')
\end{equation}

\begin{example}
  Consider once again the game graph of \figurename~\ref{fig:SP3}, but
  with a new memoryless strategy $\randomminstrategy''_p$ defined by
  $\randomminstrategy''_p(v_0) = \Dirac{v_1}$ and
  $\randomminstrategy''_p(v_1)=\delta$ such that $\delta(v_0)=1-p$ and
  $\delta(\text{\Large\smiley})=p$, where $p\in(0,1)$. Then, we can
  check that $\mValue^{\randomminstrategy''_p}(v_0)=-2$ and
  $\mValue^{\randomminstrategy''_p}(v_1)=-1$. Thus,
  $\widetilde E(v_0)=\{v_1\}$ and
  $\widetilde E(v_1)=\{v_0,\text{\Large\smiley}\}$. Not all memoryless
  deterministic strategies taken in $\widetilde {\mathcal G}$ are
  NC-strategies, since it contains the cycle $v_0v_1v_0$ of total
  weight $0$. We thus apply the construction before, using the fact
  that $d(\text{\Large\smiley})=0$, $d(v_1)=1$ and $d(v_0)=2$ (since
  the edge $(v_0,\text{\Large\smiley})$ is not present in
  $\widetilde {\mathcal G}$). Thus, $\minstrategy_1$ is defined by
  $\minstrategy_1(v_0)=v_1$ and
  $\minstrategy_1(v_1)=\text{\Large\smiley}$, and is indeed an
  NC-strategy.
\end{example}

\begin{lemma}\label{lem:NC-strategy}
  Strategy $\minstrategy_1$ is an NC-strategy, i.e.~all cycles of
  $\widetilde {\mathcal G}$ conforming with $\minstrategy_1$ have a
  negative total weight.
\end{lemma}
\begin{proof}
  Let $v_1v_2\cdots v_kv_1$ be a cycle of $\widetilde{\mathcal G}$
  that conforms to~$\minstrategy_1$, with $v_1$ a vertex of minimal
  distance~$d(v_1)$ among the ones of the cycle. We can choose~$v_1$
  such that it belongs to $\MinPl$: otherwise, this would contradict
  the attractor computation in $\widetilde {\mathcal G}$. By
  Lemma~\ref{lem:non-positive}$(ii)$, its total weight is
  non-positive. Suppose that it is $0$. Then, in the proof of
  Lemma~\ref{lem:non-positive}$(ii)$, all inequalities
  $\mValue^\randomminstrategy(v_i)\geq
  \edgeweights_i+\mValue^\randomminstrategy(v_{i+1})$ are indeed
  equalities. In particular,
  $\mValue^\randomminstrategy(v_1)=
  \edgeweights_1+\mValue^\randomminstrategy(v_{2})$. Since
  $v_2\in \widetilde E(v_1)$, \eqref{eq:final-inequality} ensures that
  all successors $v'\in \support{\randomminstrategy(v_1)}$,
  $\mValue^\randomminstrategy(v_1)=
  \edgeweights(v_1,v')+\mValue^\randomminstrategy(v')$. Since $v_1$
  has minimal distance among all vertices of the cycle, it exists
  $v'\in \widetilde E(v_1)$ such that $d(v')=d(v_1)-1$. But
  $d(v_2)\geq d(v_1)>d(v')$, which contradicts the choice of $v_2$ for
  $\minstrategy_1(v_1)$ in~\eqref{eq:sigma1}.
\end{proof}

\iffalse
\begin{example}\todo{Change the example to a more interesting one
    where we use the distance}
  Let the game represented by Figure~\ref{fig:SP2} where we change
  some weights: $w(v_1, v_2) = 0$, $w(v_2, v_3) = 0$ and
  $w(v_3, v_1) = 0$. We consider the memoryless strategy
  $\randomminstrategy_{p}$ built in the previous section: then
  $\mValue^{\randomminstrategy_{p}}(v_0) = -10$,
  $\mValue^{\randomminstrategy_{p}}(v_1) = -10p$ and
  $\mValue^{\randomminstrategy_{p}}(v_2) =
  \mValue^{\randomminstrategy_{p}}(v_3) = -10p^2$. Moreover, distances
  are $d_{v_0} = 1$, $d_{v_1} = 2$, $d_{v_2} = 2$ and $d_{v_3} =
  1$. We compute $A_{v_1} = \{v_0\}$ and $A_{v_3} = \{v_1\}$. By
  definition of $\minstrategy_1$, we thus have
  $\minstrategy_1(v_1) = v_0$ and
  $\minstrategy_1(v_3) = v_1$. Moreover,
  $\minstrategy_2(v_1) = v_2$ and $\minstrategy_2(v_3)$ is the target.
\end{example}
\fi

\begin{proof}[Proof of Proposition~\ref{prop:sigmaM}]
  Let $\pi$ be a play conforming to $\minstrategy$, from vertex
  $v_0$. Since $\minstrategy$ is a switching strategy, it necessarily
  reaches $\finalvertices$. If $\minstrategy$ conforms with
  $\minstrategy_1$, by Lemma~\ref{lem:non-positive}$(i)$, it has a
  total-payoff $\TP(\pi)\leq
  \mValue^{\randomminstrategy}(v_0)$. Otherwise, it is obtained by a
  switch, and is thus longer than
  $\alpha=\max(0,|V|W-\mValue^{\rho}(v_0))\times |V|+1$. Then, it
  contains at least $\max(0,|V|W-\mValue^{\rho}(v_0))$ elementary
  cycles, before it switches to the attractor strategy
  $\minstrategy_2$. Once we remove the cycles, it remains a play of
  length at most $|V|$, and thus of total payoff at most $|V|W$. Since
  all cycles conforming to $\minstrategy_1$ have a total weight at
  most $-1$, by Lemma~\ref{lem:NC-strategy}, $\TP(\pi)$ is at most
  $(-1)\times \max(0,|V|W-\mValue^{\rho}(v_0)) + |V|W \leq
  \mValue^{\rho}(v_0)$.
\end{proof}

This concludes the proof of Theorem~\ref{thm:main}.

% \begin{remark}
%   In this section and article, we only considered randomised
%   strategies that are memoryless. We could also define a more general
%   upper value $\puppervalue(v)$ when we let $\MinPl$ and $\MaxPl$ play
%   unrestricted strategies (randomised and with memory). The reasoning
%   of Section~3, where we do not need memory when playing with
%   randomised strategies, \emph{a fortiori} shows that
%   $\puppervalue(v)\leq \dValue(v)$\todo{Really, what about $\MaxPl$
%     that we restrict to strategies without memory in
%     Prop~\ref{prop:target-proba1}?}. Moreover, in the current section,
%   we only used the vector of values
%   $(\mValue^\randomminstrategy(v))_{v\in \vertices}$ to define the
%   switching strategy $\minstrategy$, without using anywhere that
%   $\randomminstrategy$ is memoryless. We thus indeed showed that
%   $\dValue(v)\leq \puppervalue(v)$. In total, we have that
%   $\dValue(v)= \puppervalue(v)=\muppervalue(v)$. Notice \todo{Je me
%     demande si c'est vraiment nécessaire cette dernière phrase}
%   finally that Blackwell determinacy \cite{Mar98} implies that, for
%   unrestricted strategies, shortest-path games are determined so that
%   $\puppervalue(v) = \plowervalue(v)$.
% \end{remark}

\section{Characterisation of optimality}
\label{sec:optimality}

All shortest-path games admit an optimal deterministic strategy for
both players: however, as we have seen in
Example~\ref{ex:needs-memory}, $\MinPl$ may require memory to play
optimally. In this case, we also have seen in
Example~\ref{example:for-intro} that $\MinPl$ does not have an optimal
memoryless (randomised) strategy: he only has $\epsilon$-optimal ones,
for all $\epsilon>0$. But some shortest-path games indeed admit
optimal memoryless strategies for $\MinPl$: the strategy
$\randomminstrategy_{p}$ described in Section~\ref{sec:memory2random}
is indeed optimal in graph games not containing negative cycles, for
instance. In this final section, we characterise the shortest-path
games in which $\MinPl$ admits an optimal memoryless strategy. For
sure, $\MinPl$ does not have an optimal strategy if there is some
vertex $v$ of value $\dValue(v)=-\infty$.

\begin{assumption}
  In this last section, we therefore suppose that all shortest-path
  games are such that $\dValue(v)\neq -\infty$ for all vertices $v$.
\end{assumption}

\newcommand\F{\mathcal F}

We first recall the computations performed in
\cite{BrihayeGeeraertsHaddadMonmege2016} to compute values
$\dValue(v)$. It consists of an iterated computation, called
\emph{value iteration} based on the operator
$\F\colon (\Z\cup\{+\infty\})^\vertices \to
(\Z\cup\{+\infty\})^\vertices$ defined for all
$x=(x_v)_{v\in \vertices}\in (\Z\cup\{+\infty\})^\vertices$ and all
vertices $v\in \vertices$ by
\[\F(x)_v =
  \begin{cases}
    0& \text{if } v\in \finalvertices\\
    \min_{v'\in E(v)} (\edgeweights(v,v')+x_{v'}) & \text{if } v\in
    \minvertices\\
    \max_{v'\in E(v)} (\edgeweights(v,v')+x_{v'}) & \text{if } v\in \maxvertices
  \end{cases}\] We let $f^{(0)}_v=0$ if $v\in \finalvertices$ and
$+\infty$ otherwise. By monotony of $\F$, the sequence
$(f^{(i)}=\F^i(f^{(0)}))_{i\in \N}$ is non-increasing. It is proved to
be stationary, and convergent towards $(\dValue(v))_{v\in \vertices}$,
the smallest fixed-point of $\F$. The pseudo-polynomial complexity of
solving shortest-path games comes from the fact that this sequence may
becomes stationary after a pseudo-polynomial (and not polynomial)
number of steps: the game of \figurename~\ref{fig:SP1} is one of the
typical examples.

We introduce a new notion, being the most permissive strategy of
$\MinPl$ at each step $i\geq 0$ of the computation. It maps each
vertex $v\in\minvertices$ to the set
\[\widetilde E^{(i)}(v)=\{v'\in E(v)\mid
  \edgeweights(v,v')+f^{(i-1)}_{v'}=f^{(i)}_v\}\] of vertices that
$\MinPl$ can choose. For each such most permissive strategy
$\widetilde E^{(i)}$, we let $\widetilde\game^{(i)}$ be the game graph
where we remove all edges $(v,v')$ with $v\in \minvertices$ and
$v'\notin \widetilde E^{(i)}(v)$. This allows us to state the
following result:

\begin{proposition}%[Existence of an optimal memoryless strategy]
  \label{prop:caractOptiSansMem}
  The following assertions are equivalent:
  \begin{enumerate}
  \item\label{item:1} $\MinPl$ has an optimal memoryless deterministic
    strategy in $\game$ (for $\dValue$);
  \item\label{item:2} $\MinPl$ has an optimal memoryless (randomised)
    strategy in $\game$ (for $\muppervalue$);
  \item\label{item:3} $f^{(|V|-1)}_v=f^{(|V|)}_v=\dValue(v)$ for all
    vertices $v$ (this means that the sequence $(f^{(i)})$ is
    stationary as soon as step $|V|-1$), and $\MinPl$ can guarantee to
    reach $\finalvertices$ from all vertices in the game graph
    $\widetilde\game^{(|V|-1)}$.
  \end{enumerate}
\end{proposition}
\begin{proof}
Implication $\ref{item:1}\Rightarrow \ref{item:2}$ is trivial by
  the result of Theorem~\ref{thm:main}.

  For implication $\ref{item:3}\Rightarrow \ref{item:1}$, consider
  any memoryless deterministic strategy $\minstrategy^*$ that
  guarantees $\MinPl$ to reach $\finalvertices$ from all vertices in
  the game graph $\widetilde\game^{(|V|-1)}$. Then, for all vertices $v$, we
  show by induction on $n$, that each play $\pi$ from $v$ that reaches
  the target in at most $n$ steps, and conforming to $\minstrategy^*$,
  has a total-payoff $\TP(\pi)\leq \dValue(v)$. This is trivial for
  $n=0$. If~$\pi= v \pi'$ with $\pi'$ starting in $v$, then
  \[\TP(\pi)=\edgeweights(v,v') + \TP(\pi')\leq \edgeweights (v,v') +
    \dValue(v') = \edgeweights (v,v') + f^{(|V|-1)}_v\] If
  $v\in \maxvertices$, we have
  \[\TP(\pi)\leq \edgeweights (v,v') + f^{(|V|-1)}_v \leq
    f^{(|V|)}_v=\dValue(v)\] If $v\in \minvertices$,
  since~$v'\in \widetilde E^{(|V|-1)}(v)$,
  \[\TP(\pi)= f^{(|V|)}_v = \dValue(v)\]
  This ends the proof by induction. To conclude that $\ref{item:1}$
  holds, since $\minstrategy^*$ guarantees to reach the target, all
  plays conforming to it reach the target in less than $|V|$ steps,
  which proves that~$\dValue^{\minstrategy^*}(v)\leq \dValue(v)$,
  showing that~$\minstrategy^*$ is optimal.

    \medskip

  For implication $\ref{item:1}\Rightarrow\ref{item:3}$, consider an
  optimal deterministic memoryless strategy $\minstrategy^*$, such
  that for all $v$, $\dValue^{\minstrategy^*}(v)=\dValue(v)$.
  
  First, we show that $f^{(|V|-1)}_v=\dValue(v)$ for all vertices
  $v$. For that, consider the deterministic strategy $\maxstrategy$ of
  $\MaxPl$ defined for all finite plays $\pi$ having $n\leq |V|$
  vertices, ending in a vertex $v\in\maxvertices$, by
  $\maxstrategy (\pi) = v'$ such that
  $\edgeweights(v,v')+f_{v'}^{(|V|-1-n)}=f_{v}^{(|V|-n)}$. For longer
  finite plays, we define $\maxstrategy$ arbitrarily. Then, let $\pi$
  be the play from $v$ conforming to $\minstrategy^*$
  and~$\maxstrategy$. Since $\minstrategy^*$ ensures reaching the
  target and is memoryless deterministic, $\pi$ reaches the target in
  at most $|V|-1$ steps. Let $\pi=v_0v_1v_2\cdots v_{k-1}v_k$ with
  $k\leq |V|$. Let us show that $\TP(\pi)\geq f_v^{(|V|-1)}$. We prove
  by induction on $0\leq j\leq k$ that
  \[\sum_{i=j}^{k-1} \edgeweights(v_i,v_{i+1})\geq
    f_{v_j}^{(|V|-1-j)}\] When $j=k-1$, the result is trivial since
  the sum is \[0=f_{v_k}^{(0)}\geq f_{v_k}^{(|V|-1-(k-1))}\]
  Otherwise,
  by induction hypothesis
  \[\sum_{i=j}^{k-1} \edgeweights(v_i,v_{i+1}) \geq
    \edgeweights(v_j,v_{j+1}) +f_{v_{j+1}}^{(|V|-1-(j+1))}\] If
  $v_j\in \maxvertices$, $v_{j+1}$ is chosen by $\maxstrategy$ so that
  \[\edgeweights(v_j,v_{j+1})+f_{v_{j+1}}^{(|V|-1-(j+1))}=
    f_{v_j}^{(|V|-1-j)}\] If $v\in \minvertices$, by definition of
  $\F$,
  \[\edgeweights(v_j,v_{j+1})+f_{v_{j+1}}^{(|V|-1-(j+1))}\geq
    f_{v_j}^{(|V|-1-j)}\] We can conclude in all cases, so that
  $f^{(|V|-1)}_v=\dValue(v)$ for all vertices $v$.

  Then, we show that $\MinPl$ can guarantee to reach $\finalvertices$
  from all vertices in the game graph~$\widetilde\game^{(|V|-1)}$. Let us
  suppose that this is not the case. Then, there exists a set
  $\vertices'$ of vertices in which $\MaxPl$ can guarantee to keep
  $\MinPl$ for ever, in the game $\widetilde\game^{(|V|-1)}$: for all
  $v'\in\vertices'\cap\minvertices$,
  $\widetilde E^{(|V|-1)}(v') \subseteq \vertices'$, and for all
  $v'\in\vertices'\cap\maxvertices$,
  $E(v)\cap \vertices'\neq \emptyset$. Since~$\minstrategy^*$
  guarantees to reach the target, there exists
  $v\in \vertices'\cap\minvertices$ such that
  $\minstrategy^*(v)=v'\notin\vertices'$: then
  $\edgeweights(v,v')+\dValue(v') > \dValue(v)$ (here we use that
  $\dValue(v)=f_v^{(|V|-1)} = f_v^{(|V|)}$). Consider an optimal
  deterministic memoryless strategy $\maxstrategy^*$ of
  $\MaxPl$ in $\game$. Then, the play $\pi$ from $v$ conforming to
  $\minstrategy^*$ and $\maxstrategy^*$ starts by taking the edge
  $(v,v')$ and continues with a play $\pi'$. By optimality, we know
  that $\TP(\pi)=\dValue(v)$ and $\TP(\pi')=\dValue(v')$. However,
  \[\TP(\pi)=\edgeweights(v,v')+\TP(\pi') =
    \edgeweights(v,v')+\dValue(v') > \dValue(v)\] which raises a
  contradiction.

  \medskip

  We finish the proof by showing
  $\ref{item:2}\Rightarrow \ref{item:1}$. For that, consider an
  optimal memoryless strategy~$\randomminstrategy^*$ for
  $\muppervalue$. By following the construction of
  Section~\ref{sec:random2memory}, we build a memoryless deterministic
  strategy $\minstrategy_1$. Lemma~\ref{lem:NC-strategy} ensures that
  $\minstrategy_1$ is an NC-strategy so that every cycle conforming
  to $\minstrategy_1$ has a negative total weight. Let us show that
  such a negative cycle cannot exist, which will ensure that all plays
  conforming to $\minstrategy_1$ reach the target, and thus the
  optimality of $\minstrategy_1$. Suppose that a cycle
  $v_1v_2\cdots v_kv_1$ conforms to $\minstrategy_1$.  By following
  the notations of the proof of Lemma~\ref{lem:non-positive}$(ii)$, we
  suppose that $v_1$ is a vertex of minimal distance $d(v_1)$ to the
  target, and that it is owned by $\minvertices$.  Note that such a
  vertex exists, otherwise only $\MaxPl$ has the minimal distance
  vertices on the cycle and that contradicts the attractor
  computation.  By minimality of $d(v_1)$ among the vertices of the
  cycle, $d(v_2) \geq d_{v_1}$. Moreover, by the attractor
  computation, there exists $u \in E(v_1)$ such that
  $d(u)=d(v_1)-1 < d(v_1)$. By definition of~$\minstrategy_1$, we know
  for sure that $u\notin \widetilde E(v_1)$, so that
  \[\edgeweights(v_1, u) + \mValue^{\randomminstrategy^*}(u) >
    \edgeweights (v_1,v_2) + \mValue^{\randomminstrategy^*}(v_2)\]
  By~\eqref{eq:final-inequality}, we know that in this case
  \[\mValue^{\randomminstrategy^*}(v_{1})> \edgeweights(v_1,v_2)
    +\mValue^{\randomminstrategy^*}(v_{2})\] By optimality
  of~$\randomminstrategy^*$, this rewrites in
  \[\muppervalue(v_{1})> \edgeweights(v_1,v_2)
    +\muppervalue(v_{2})\] By Theorem~\ref{thm:main}, this also
  rewrites in
  \[\dValue(v_{1})> \edgeweights(v_1,v_2)+\dValue(v_{2})\geq
    \F\big((\dValue(v))_{v\in \vertices}\big)(v_1)\] (since
  $v_1 \in \minvertices$): this contradicts the fact that the vector
  $(\dValue(v))_{v\in \vertices}$ is a fixed-point of~$\F$.
\end{proof}

This characterisation of the existence of optimal memoryless strategy
is testable in polynomial time since it is enough to compute vectors
$f^{(|V|-1)}$ and $f^{(|V|)}$, check their equality, compute the sets
$\widetilde E^{(|V|-1)}(v)$ (this can be done while computing
$f^{(|V|)}$) and check whether $\MinPl$ can guarantee reaching the
target in $\widetilde\game^{(|V|-1)}$ by an attractor computation. The
proof of implication $\ref{item:3}\Rightarrow \ref{item:1}$ is
constructive and actually allows one to build an optimal memoryless
deterministic strategy when it exists.

\section{Discussion}
\label{sec:conclusion}

This article studies the tradeoff between memoryless and deterministic
strategies, showing that $\MinPl$ guarantees the same value when
restricted to these two kinds of strategies. We also studied the
existence of optimal memoryless strategies, which turns out to be
equivalent to the existence of optimal memoryless deterministic
strategies, and testable in polynomial time.

We could also define a more general lower and upper values
$\plowervalue(v)$/$\puppervalue(v)$ when we let $\MinPl$ and $\MaxPl$
play unrestricted strategies (randomised and with memory). The
Blackwell determinacy results \cite{Mar98} implies that, for such
unrestricted strategies, shortest-path games are still determined so
that $\puppervalue(v) = \plowervalue(v)=\pValue(v)$. The reasoning of
Section~\ref{sec:random2memory} only used the vector of values
$(\mValue^\randomminstrategy(v))_{v\in \vertices}$ to define the
deterministic switching strategy $\minstrategy$, without using
anywhere that $\randomminstrategy$ is memoryless. We thus indeed
showed that $\dValue(v)\leq \pValue(v)$. However, the proof of
Section~\ref{sec:memory2random} is not directly translatable if we
allow $\MinPl$ to use memory and randomisation. In particular, we know
nothing anymore about how $\MaxPl$ can react, which may break the
result of Proposition~\ref{prop:target-proba1}. We leave this further
study for future work.


\begin{thebibliography}{10}

\bibitem{BaierKatoen}
Christel Baier and Joost{-}Pieter Katoen.
\newblock {\em Principles of model checking}.
\newblock {MIT} Press, 2008.

\bibitem{BertsekasTsitsiklis91}
Dimitri~P. Bertsekas and John~N. Tsitsiklis.
\newblock An analysis of stochastic shortest path problems.
\newblock {\em Math. Oper. Res.}, 16(3):580--595, 1991.

\bibitem{BrihayeGeeraertsHaddadMonmege2016}
Thomas Brihaye, Gilles Geeraerts, Axel Haddad, and Benjamin Monmege.
\newblock Pseudopolynomial iterative algorithm to solve total-payoff games and
  min-cost reachability games.
\newblock {\em Acta Informatica}, 54, 07 2016.

\bibitem{BGHM17}
Thomas Brihaye, Gilles Geeraerts, Axel Haddad, and Benjamin Monmege.
\newblock Pseudopolynomial iterative algorithm to solve total-payoff games and
  min-cost reachability games.
\newblock {\em Acta Informatica}, 54(1):85--125, February 2017.
\newblock \href {https://doi.org/10.1007/s00236-016-0276-z}
  {\path{doi:10.1007/s00236-016-0276-z}}.

\bibitem{ChaAlfHen04}
Krishnendu Chatterjee, Luca~de Alfaro, and Thomas~A. Henzinger.
\newblock Trading memory for randomness.
\newblock In {\em Proceedings of the The Quantitative Evaluation of Systems,
  First International Conference}, QEST '04, pages 206--217, Washington, DC,
  USA, 2004. IEEE Computer Society.

\bibitem{ChaHen05}
Krishnendu Chatterjee, Thomas~A. Henzinger, and Marcin Jurdzi{\'n}ski.
\newblock Mean-payoff parity games.
\newblock In {\em Proceedings of the 20th Annual Symposium on Logic in Computer
  Science (LICS'05)}, pages 178--187. {IEEE} Computer Society Press, 2005.

\bibitem{ChaHenPra208}
Krishnendu Chatterjee, Thomas~A. Henzinger, and Vinayak~S. Prabhu.
\newblock Trading infinite memory for uniform randomness in timed games.
\newblock In {\em Hybrid Systems: Computation and Control, 11th International
  Workshop, {HSCC} 2008, St. Louis, MO, USA, April 22-24, 2008. Proceedings},
  pages 87--100, 2008.
\newblock \href {https://doi.org/10.1007/978-3-540-78929-1\_7}
  {\path{doi:10.1007/978-3-540-78929-1\_7}}.

\bibitem{ChaRan14}
Krishnendu Chatterjee, Mickael Randour, and Jean-Fran{\c c}ois Raskin.
\newblock Strategy synthesis for multi-dimensional quantitative objectives.
\newblock {\em Acta Informatica}, 51:129--163, 2014.
\newblock \href {https://doi.org/https://doi.org/10.1007/s00236-013-0182-6}
  {\path{doi:https://doi.org/10.1007/s00236-013-0182-6}}.

\bibitem{GimZie04}
Hugo Gimbert and Wies{\l}aw Zielonka.
\newblock When can you play positionally?
\newblock In {\em Proceedings of the 29th International Conference on
  Mathematical Foundations of Computer Science (MFCS'04)}, volume 3153 of {\em
  Lecture Notes in Computer Science}, pages 686--698. Springer, 2004.

\bibitem{GraTho02}
Erich Gr{\"a}del, Wolfgang Thomas, and Thomas Wilke.
\newblock {\em Automata, Logics, and Infinite Games: A Guide to Current
  Research}, volume 2500 of {\em Lecture Notes in Computer Science}.
\newblock Springer, 2002.

\bibitem{KhaBor08}
Leonid Khachiyan, Endre Boros, Konrad Borys, Khaled Elbassioni, Vladimir
  Gurvich, Gabor Rudolf, and Jihui Zhao.
\newblock On short paths interdiction problems: Total and node-wise limited
  interdiction.
\newblock {\em Theory of Computing Systems}, 43:204--233, 2008.

\bibitem{Mar98}
Donald~A. Martin.
\newblock The determinacy of {B}lackwell games.
\newblock {\em The Journal of Symbolic Logic}, 63(4):1565--1581, 1998.

\bibitem{Nas50}
John~F. Nash.
\newblock Equilibrium points in n-person games.
\newblock {\em Proceedings of the National Academy of Sciences of the United
  States of America}, 36(1):48--49, 1950.

\end{thebibliography}
\end{document}